\documentclass[10pt]{elsarticle}

\usepackage[utf8]{inputenc}
\usepackage{float}
\usepackage{verbatim}
\usepackage{mwe}
\usepackage{charter}
\usepackage{afterpage}
\usepackage{amsmath}
\usepackage[title]{appendix}
\usepackage{ragged2e}
\usepackage{etoolbox}
\usepackage{fancyhdr}
\usepackage{booktabs}
\usepackage{arydshln}
\usepackage{enumitem}
\usepackage{mathtools}

\usepackage{subfiles}
\usepackage[compact]{titlesec}
\usepackage{blindtext}
\usepackage{stfloats}
\usepackage{lipsum}

\usepackage{physics}
\usepackage[english]{babel}
\usepackage{natbib}
\usepackage{multirow}

\usepackage{amssymb}
\usepackage{nomencl}
\usepackage{titlesec}
\usepackage{latexsym}
\usepackage{amsfonts}
\usepackage{times}
\usepackage{url}
\usepackage{rotating}
\DeclareGraphicsExtensions{.jpg,.pdf,.png,.eps}
\usepackage{fancyhdr}
\usepackage{anysize}
\usepackage{bm}
\usepackage{bbm}
\usepackage{stmaryrd}
\usepackage{hyperref}
\usepackage{subcaption}
\usepackage{caption}
\usepackage{amsmath}
\usepackage{subcaption}
\usepackage{graphicx}

\usepackage{parskip}
\usepackage{algpseudocode,algorithm,algorithmicx}
\usepackage{diagbox}
\usepackage[table]{xcolor}

\usepackage{caption}

\DeclareMathOperator*{\argmin}{argmin}

\begin{document} 
\begin{frontmatter} 
	\title{Towards Generative Design Using Optimal Transport for Shape Exploration and Solution Field Interpolation} 
	\author[1]{Sergio Torregrosa\corref{cor1}} 
	\ead{sergio.torregrosa_jordan@ensam.eu} 
	\author[2,3]{David Muñoz} 
	\ead{david.munoz_pellicer@ensam.eu} 
	\author[2]{Hector Navarro} 
	\ead{hector.navarro_garcia@ensam.eu}
	\author[4]{Charbel Farhat}
	\ead{cfarhat@stanford.edu}
	\author[5]{Francisco Chinesta}
	\ead{francisco.chinesta@ensam.eu}
	
	\cortext[cor1]{Corresponding author}
	
	\address[1]{ENSAM Institute of Technology, PIMM, UMR CNRS 8006, ESI/Keysight Chair, 151 Boulevard de l’Hopital, 75013 Paris, France}
	\address[2]{ENSAM Institute of Technology, PIMM, UMR CNRS 8006, 151 Boulevard de l’Hopital, 75013 Paris, France}
	\address[3]{Ecole Centrale de Nantes, GeM, UMR CNRS 6183, 1 Rue de la Noë, 44321 Nantes, France}        	
	\address[4]{Stanford University, Department of Aeronautics and Astronautics and Institute for Computational and Mathematical 
	Engineering, Stanford, CA 94305-4035, USA}
	\address[5]{CNRS@CREATE Ltd., 1 CREATE Way, 08-01, CREATE Tower, 138602 Singapore} 
	\begin{abstract} 
	Generative Design (GD) combines artificial intelligence (AI), physics-based modeling, and multi-objective optimization to 
	autonomously explore and refine engineering designs. Despite its promise in aerospace, automotive, and other high-performance 
	applications, current GD methods face critical challenges: AI approaches require large datasets and often struggle to 
	generalize; topology optimization is computationally intensive and difficult to extend to multiphysics problems; and model 
	order reduction for evolving geometries remains underdeveloped. To address these challenges, we introduce a unified, 
	structure-preserving framework for GD based on optimal transport (OT), enabling simultaneous interpolation of complex 
	geometries and their associated physical solution fields across evolving design spaces, even with non-matching meshes and 
	substantial shape changes. This capability leverages Gaussian splatting to provide a continuous, mesh-independent 
	representation of the solution and Wasserstein barycenters to enable smooth, mathematically ''mass''-preserving blending of 
	geometries, offering a major advance over surrogate models tied to static meshes. Our framework efficiently interpolates 
	positive scalar fields across arbitrarily shaped, evolving geometries without requiring identical mesh topology or 
	dimensionality. OT also naturally preserves localized physical features -- such as stress concentrations or sharp gradients 
	-- by conserving the spatial distribution of quantities, interpreted as ''mass'' in a mathematical sense, rather than 
	averaging them, avoiding artificial smoothing. Preliminary extensions to signed and vector fields are presented. 
	Representative test cases demonstrate enhanced efficiency, adaptability, and physical fidelity, establishing a foundation for 
	future foundation-model-powered generative design workflows.
	\end{abstract}
	\begin{keyword} {Generative Design; Optimal Transport; Gaussian splatting; Interpolation; Parametric Metamodeling; Regression; 
	                 Structure-Preserving; Localized Features}
\end{keyword}
\end{frontmatter} 


\section{Introduction} 
\label{sec:INTRO}

Generative design (GD) leverages computational intelligence to autonomously explore, evaluate, and refine design solutions, 
integrating artificial intelligence (AI), physics-based modeling, and multi-objective optimization. By combining these capabilities,
GD enables rapid discovery of innovative, high-performance designs that challenge conventional notions of efficiency, functionality,
and manufacturability. Modern GD approaches employ diverse strategies -- including topology optimization, deep learning, reinforcement
learning, and procedural generation -- to generate designs that satisfy geometric, structural, material, or performance criteria
\cite{zhu2016topology, yang2018microstructural, chen2020airfoil, heyrani2021range}. While such methods have shown success in
applications spanning aerospace, automotive, and architecture \cite{autodesk, stackpole, ackerman}, they still face major challenges
related to generalization, computational cost, and the seamless integration of physical constraints \cite{vaneker2020design,
parrott2023multidisciplinary, wang2023overview}.

A useful way to classify GD methods is by how they leverage available information: (i) \textit{algorithmic approaches} rely on 
rule-based grammars and procedural heuristics \cite{chakrabarti2011computer}; (ii) \textit{physics-based approaches}, such as topology
optimization, employ governing equations to distribute material optimally but remain computationally demanding and difficult to extend
to multiphysics settings; and (iii) \textit{data-driven and hybrid physics-informed approaches} combine machine learning and 
projection-based model order reduction to enhance adaptability and efficiency \cite{benner2021model, choi2020gradient, 
boncoraglio2022piecewise}. Nevertheless, despite these methodological advances, most AI-based GD methods still rely on massive 
datasets, while state-of-the-art projection-based model order reduction techniques remain confined to topologically invariant meshes 
\cite{little2024projection} -- a serious limitation for realistic GD tasks involving shape evolution and adaptive meshing.

To realize a broader, data-driven GD framework -- one potentially suitable for future integration with foundation models -- we focus
on developing a methodology to \textbf{learn and infer the mapping} between design parameters, complex evolving geometries, and their 
optimal physical solutions. Instead of computationally expensive direct optimization, our framework accelerates the design cycle 
through efficient, data-driven inference of the full solution field.

As an essential step toward this vision, we develop a unified, structure-preserving computational methodology based on optimal 
transport (OT) theory for efficient and accurate interpolation and regression of physical fields. While OT offers superior 
interpolation properties -- capturing spatial displacements and what we term ``localized features'' more naturally than Euclidean 
methods -- it has historically been computationally prohibitive for real-time applications \cite{burkard2012assignment, weed2019sharp}.
Throughout this paper, ``localized features'' refers to small-scale or region-specific phenomena such as stress concentrations, 
cracks, sharp gradients, or vortices. Building upon the parametric OT-based surrogate model of \cite{torregrosa2022surrogate}, we 
extend it to support GD by integrating it with a Wasserstein barycenter formulation that enables interpolation across arbitrarily 
shaped, out-of-sample geometries. This development significantly enhances generalization and efficiency in the generative process 
while preserving physical structure.

The resulting methodology performs interpolation and regression of parametric solutions associated with new geometries without
requiring identical mesh dimensionality, a property that is essential for generative design workflows where
geometries evolve continuously and may undergo topological changes during optimization and exploration. While our formulation is
developed for positive scalar fields, we also present preliminary extensions to signed and vector fields. In all cases, the approach
maintains structure preservation by design, ensuring that the reconstructed quantities retain the physical form and sign of the
original fields.

\begin{figure}[h!] 
	\centering 
	\includegraphics[width=0.8\textwidth]{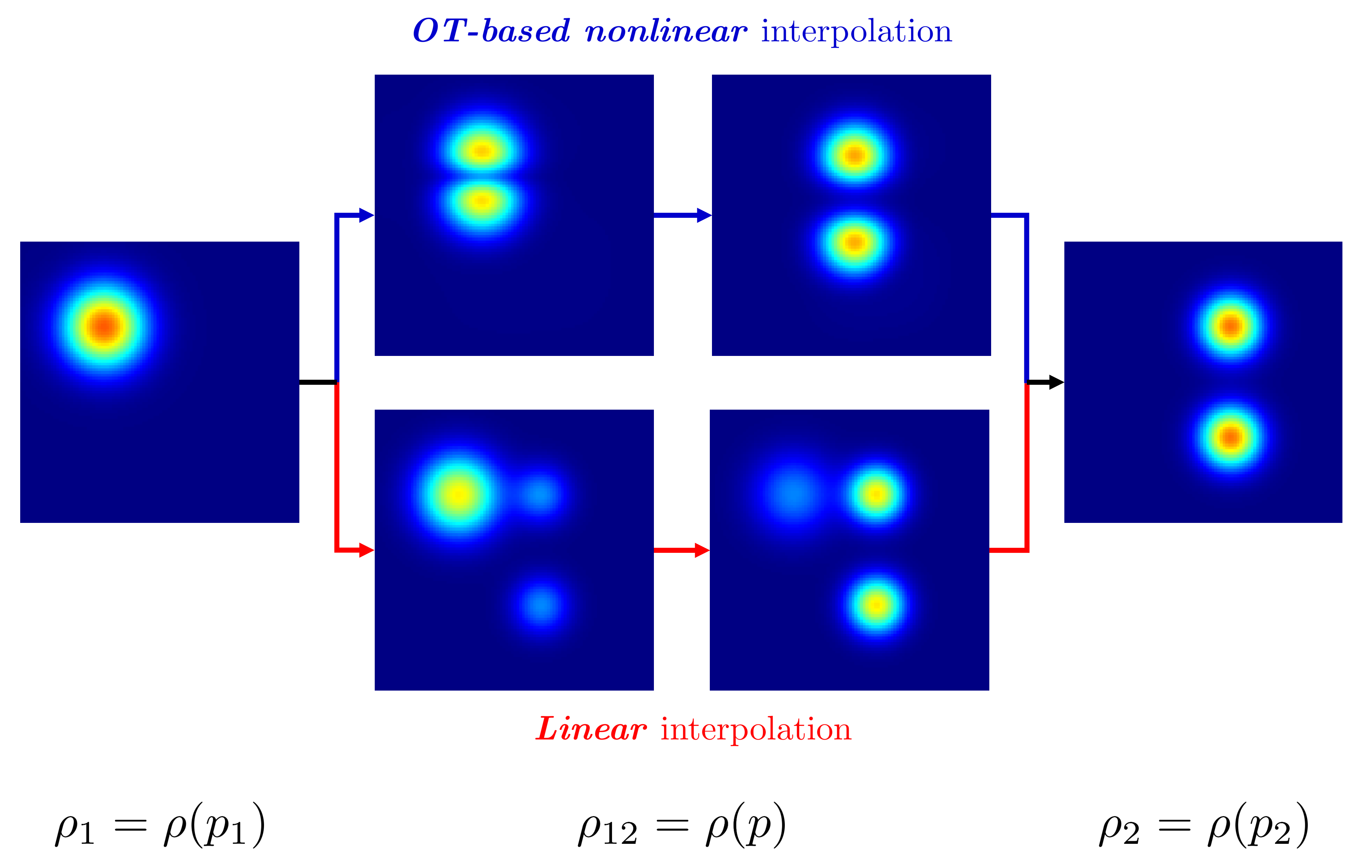} 
	\caption{Comparison of two approximation methods for a scalar field $\rho(t)$ varying with parameter $t$, transitioning from 
	$\rho_1$ (left) to $\rho_2$ (right): OT-based interpolation (top middle) preserves localized features, while conventional 
	linear interpolation $\rho(t) = (1-t)\rho_1 + t\rho_2$ (bottom middle) introduces spurious artifacts.}
	\label{fig:OTvsLinear} 
\end{figure}

Importantly, the proposed OT-based interpolation framework is particularly effective for fields characterized by localized features
that occur in limited areas of a structure or flow but can significantly influence the overall behavior. Conventional Euclidean 
interpolation of such fields often produces spurious artifacts due to its inability to capture feature translation or deformation. 
In contrast, OT-based interpolation naturally preserves these localized features by transporting ``mass'' -- in the mathematical 
sense of a measure representing the distribution of a quantity -- or, more generally, any quantity of interest such as energy, 
temperature, or pressure, in a physically meaningful way, as illustrated in Figure \ref{fig:OTvsLinear}.

This work presents a unified framework for GD that addresses the shortcomings of current approaches by combining OT, advanced field
decomposition, and geometric blending. The core contributions of this paper that enable \textbf{simultaneous and accurate exploration 
of both design parameters and evolving geometries} are summarized as follows:

\begin{itemize}
	\item \textbf{Unified Structure-Preserving Framework}: We establish a computationally efficient OT-based methodology for the
		parametric interpolation and regression of field solutions across complex, non-matching geometries.
	\item \textbf{Geometry-Agnostic Solution Interpolation}: We achieve the ability to operate across arbitrarily shaped and
		evolving domains by utilizing Gaussian splatting to transform both the physical solution and geometry into a unified,
		mesh-independent particle representation.
	\item \textbf{Data-Efficient Geometry Blending}: We integrate a Wasserstein barycenter formulation into the OT framework,
		allowing for the stable and accurate interpolation of novel geometries from a sparse set of sampled designs, thereby
		supporting efficient shape exploration.
	\item \textbf{Robust Feature Preservation}: The use of OT naturally ensures the preservation of localized physical features,
		such as stress concentrations or sharp gradients, which are often lost or corrupted by traditional Euclidean
		interpolation methods.
\end{itemize}

The remainder of this paper is organized as follows. Section \ref{sec:BACKPLUS} reviews the foundational OT-based surrogate modeling
technique introduced in \cite{torregrosa2022surrogate} and details its extension for parametric positive scalar fields using a 
Gaussian splatting-based particle representation. Section \ref{sec:GENINTERP} extends this approach to handle geometry exploration 
and the interpolation of solutions across changing shapes by incorporating the Wasserstein barycenter concept. Section \ref{sec:EXT} 
outlines the preliminary extensions to arbitrarily signed and vector fields. Finally, Section \ref{sec:APP} provides representative 
application examples and performance evaluation in the context of a parametric heat transfer problem, and Section \ref{sec:CONC} 
concludes the paper.

\section{Interpolation of Parametric Positive Scalar Field Solution Snapshots Using Optimal Transport}
\label{sec:BACKPLUS}

This section presents the OT-based interpolation methodology from \cite{torregrosa2022surrogate}, including its nomenclature and 
mathematical notation. Leveraging OT principles, the methodology encompasses defining a cost function, formulating an optimal 
transport problem, computing the optimal transport plan or map, and interpolating between distributions. It is organized into two
primary stages, as follows:
\begin{enumerate}
	\item \textbf{Offline Stage}: In this preparatory phase, precomputed parametric field solution snapshots are stored in a 
		database $\mathcal{DB}$. Each snapshot to be interpolated is decomposed into a sum of Gaussian functions -- an 
		approach known as Gaussian splatting \cite{bao20253d} -- all sharing the same standard deviation (bandwidth) but 
		differing in their centers. This decomposition transforms field solutions into point clouds, where each point 
		corresponds to the center of a Gaussian function. Throughout this paper, the term ``point’’ refers to both the 
		Gaussian distribution and, more specifically, its center (with a predetermined standard deviation). For visual 
		clarity, all figures represent Gaussian functions by their centers. These points, or ``particles,’’ deliberately 
		evoke the terminology of Smoothed Particle Hydrodynamics (SPH) \cite{liu2010smoothed}. This particle-based 
		representation enables efficient manipulation of the data, providing a robust and accurate foundation for 
		interpolation.
	\item \textbf{Online Stage}: During this stage, the goal is to estimate the field solution at out-of-sample parameter vectors.
		Leveraging the particle representations from the offline stage, interpolation is performed using data from 
		$\mathcal{DB}$. This approach ensures that new solutions can be rapidly and accurately inferred without the need 
		for exhaustive computations.
\end{enumerate}

We emphasize the importance of positive (or non-negative) scalar fields, consistent with OT theory, which seeks the most efficient way
to transport ``mass'' -- not necessarily physical, but in the mathematical sense of a measure representing any scalar quantity -- 
between distributions. In this deterministic context, these distributions describe how the quantity is allocated across a space, 
ensuring they are inherently positive scalar fields.

Examples of positive scalar fields in solid mechanics include the norms of displacement, velocity, or acceleration fields, as well as von 
Mises stress, effective stress, and strain energy density. In fluid mechanics, examples include absolute pressure, density, temperature, 
concentration (in fluid mixtures), specific internal energy, enthalpy, and vorticity magnitude.

Additionally, we remind the reader that OT-based interpolation methods function by interpolating a parametric field along the transport 
map linking the source and target (discrete) spaces. When the dimensions of both spaces are equal, minimizing the total transport
cost -- while ensuring that each mass in the source corresponds to a unique mass in the target -- results in a one-to-one mapping or 
pairing between the distributions. In this scenario, which is the focus of this work, the optimal transport problem simplifies naturally 
to an optimal matching problem.

In Section \ref{sec:EXT}, we introduce a preliminary extension of the OT-based methodology outlined in this paper, enabling it to
accommodate arbitrarily signed and vector fields.

\subsection{Particle-Based Representation, Optimal Matching, and Regressor Training}
\label{sec:OFFLINE}

Let $\mathcal{D} \subset \mathbb{R}^Q$ denote the parameter domain of interest, and let us represent a set of $P$ sampled parameter
vectors as $\{\bm{\theta}^p \in \mathcal{D}\}_{p=1}^{P}$. Correspondingly, we define the set of positive scalar field solution 
snapshots as $\{\psi^p = \psi(\bm{X}, \bm{\theta}^p): \Omega \subset \mathbb{R}^d \to \mathbb{R}^+\}_{p=1}^{P}$, where $\Omega$ 
is the $d$-dimensional physical domain of interest and $\mathbf{X} = (x, y) \in \mathbb{R}^d$ denotes the position vector of a 
point within this domain. In this subsection, we assume that $\Omega$ is independent of $\bm{\theta}$ and that $\mathcal{D}$ does
not contain geometric parameters.

\begin{figure}[H] 
	\centering 
	\includegraphics[width=0.5\textwidth]{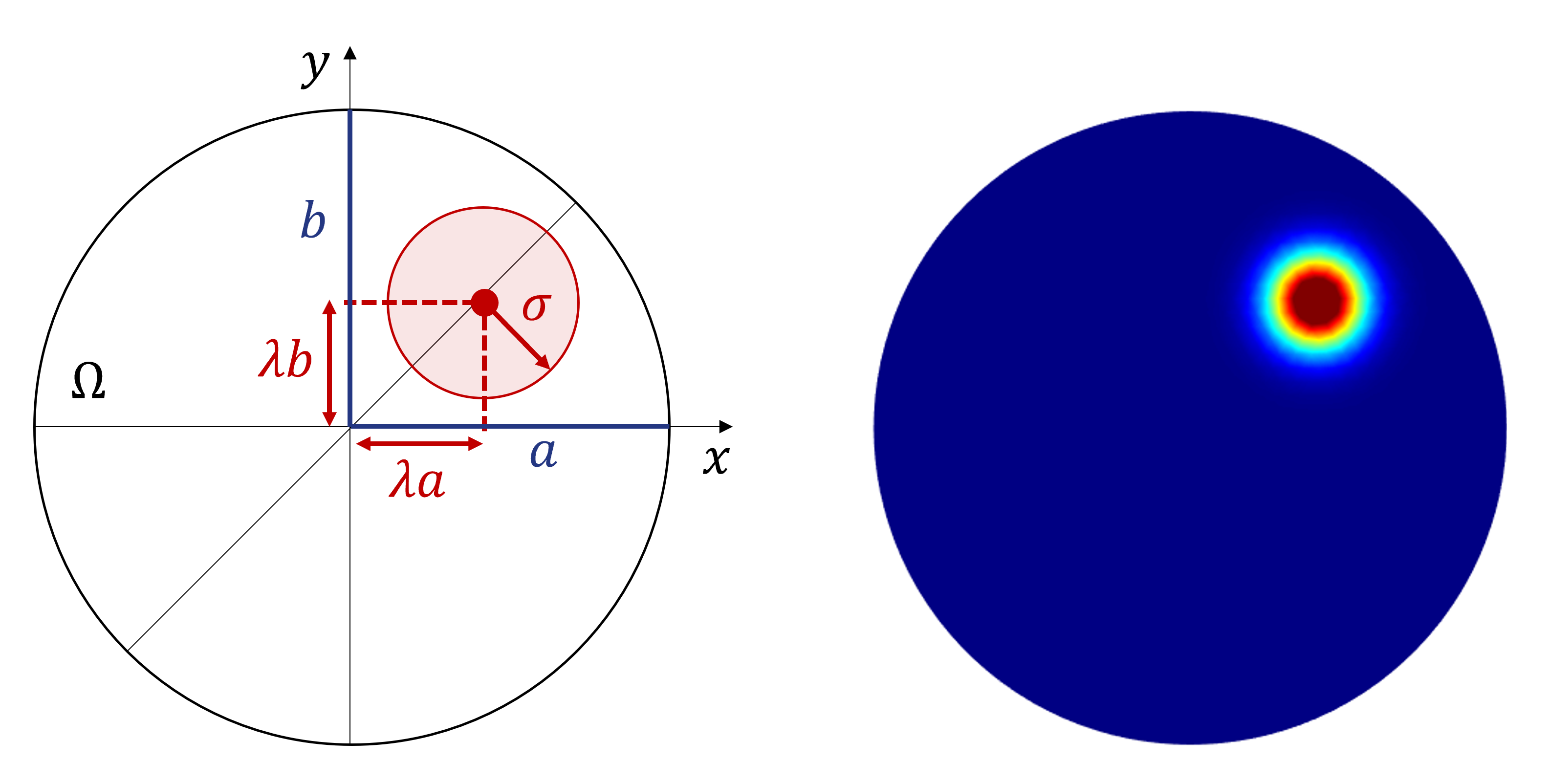} 
	\caption{Two-dimensional ellipsoidal domain, with a circular instance shown (left); positive scalar field solution 
	snapshot $\psi^p: \Omega \subset \mathbb{R}^{2} \rightarrow \mathbb{R}^+$ (right).}
	\label{fig:ToyProblem} 
\end{figure}

To illustrate the OT-based interpolation methodology as we present it here and in the following subsections, we consider a 
two-dimensional $(d=2)$ ellipsoidal domain characterized by its semi-major axis $a$ and semi-minor axis $b$, chosen for simplicity 
of presentation. We first examine the special case where $b = a$, which corresponds to a circular domain 
(see Figure \ref{fig:ToyProblem}). We define the problem of interpolating two or more parametric positive scalar fields that exhibit 
localization within $\Omega$ as problem $\mathcal{P}$.

Next, we introduce a Gaussian function with standard deviation $\sigma$ and mean positioned at $(\lambda a, \lambda b)$, where 
$\lambda \in [-1, 1]$ represents a fraction of the semi-major axis length $a$ and semi-minor axis length $b$ 
(see Figure \ref{fig:ToyProblem}).

The offline stage of the OT-based interpolation methodology introduced in \cite{torregrosa2022surrogate} approximates precomputed
positive scalar field solution snapshots (denoted as $\psi^p$) by representing them as sums of identical Gaussian functions. It 
also lays the groundwork for similar approximations of on-demand solutions. The stage consists of the following steps:

\begin{enumerate} 
	\item \textbf{Preprocessing}: Normalize the solution snapshots $\psi^{p}$ to obtain distributions with unit integral 
		values, as follows
		\[ \rho^{p} = \frac{\psi^{p}}{I^{p}}, \quad \text{where} \quad I^{p} = \int_{\Omega} \psi^{p} \, d\Omega \] 
	\item \textbf{Particle Decomposition}: Approximate each distribution as a superposition of $N_{s}$ identical two-dimensional 
		Gaussian functions, referred to as particles (see Figure \ref{fig:Particles}, left and right). Each particle has a 
		fixed standard deviation $\sigma_{s}$ and an integral equal to $1/N_{s}$, where the subscript $s$ denotes quantities 
		associated with solution snapshots. Treat both the number of particles $N_{s}$ and the standard 
		deviation $\sigma_{s}$ as hyperparameters to be tuned during the offline stage. Therefore, express each approximation 
		as
		\begin{equation} 
			\label{eq:DING}
			\rho^p (\mathbf{X}) \approx \hat{\rho}^p(\mathbf{X}) = \sum_{n=1}^{N_{s}} G_{\bm{\mu}^p_n, \sigma_s} 
			(\mathbf{X}), \quad \text{where} \quad G_{\bm{\mu}^p_n, \sigma_{s}}(\mathbf{X}) = \frac{1}{N_{s} 
			\sigma_{s}^2 2\pi} \exp\left( - \frac{(\mathbf{X} - \bm{\mu}^p_n)^2}{2\sigma_{s}^2} \right)
		\end{equation}
		where the ``coordinates'' $\mu^p_{n,x}$ and $\mu^p_{n,y}$ of the two-dimensional vector $\bm{\mu}^p_n$ represent
		the positions of the $n$-th Gaussian function's mean along the $x$ and $y$ axes, respectively. Thus, the matrix 
		$\bm{\mu}^p \in \mathbb{R}^{N_{s} \times 2}$ containing the $x$ and $y$ coordinates of all particles can be
		written as
		\[\bm{\mu}^p = \begin{bmatrix} \bm{\mu}^p_1 \\ \vdots \\ \bm{\mu}^p_n \\ \vdots \\ \bm{\mu}^p_{N_s} \end{bmatrix}
			= \begin{bmatrix} \mu^p_{1,x} & \mu^p_{1,y} \\ \vdots & \vdots \\ \mu^p_{n,x} & \mu^p_{n,y} \\ 
				\vdots & \vdots \\ \mu^p_{N_s,x} & \mu^p_{N_s,y} \end{bmatrix}\] 
	\item \textbf{Optimization}: To determine the particle positions, solve $P$ optimization problems, one for each 
		distribution, minimizing the error between the original and the Gaussian-reconstructed distributions
		\begin{equation}
			\label{eq:minpb1} 
			\argmin_{\bm{\mu}^p} \frac{1}{2} \left\| \rho^p - \hat{\rho}^p \right\|_{2}^{2} = \argmin_{\bm{\mu}^p} 
			\frac{1}{2} \left( \sum_{m=1}^{M} \left( \rho^p (\mathbf{X}_{m}) - \sum_{n=1}^{N_{s}} G_{\bm{\mu}^p_n, 
			\sigma_{s}}(\mathbf{X}_{m}) \right)^2 \right), \quad p = 1,~\cdots,~P
		\end{equation}
		where $M$ is the number of points in $\Omega$ chosen for representing the distribution $\rho^{p}$. The evaluation 
		points $\mathbf{X}_{m}$ are selected on a uniform grid sufficiently fine to capture all relevant features of the field.

		\begin{figure}[H] 
			\centering 
			\includegraphics[width=0.7\textwidth]{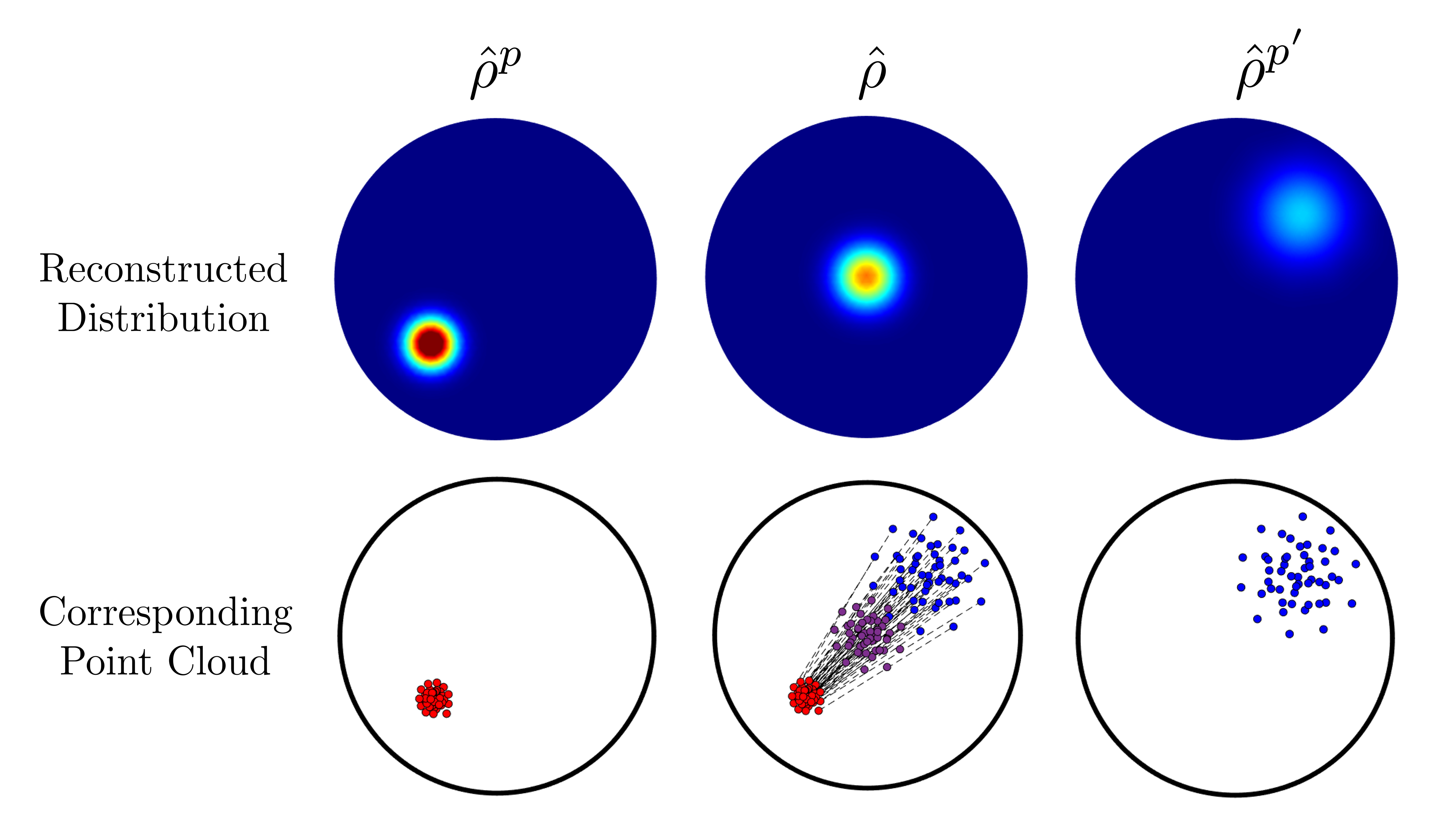} 
			\captionsetup{width=0.8\linewidth}
			\caption {Distributions and corresponding point clouds: (a) original distribution 
			$\hat{\rho}^p$ with red point cloud; (b) partially displaced violet point cloud, its reconstructed 
			distribution $\hat{\rho}$, and optimal matching between point clouds (black lines); (c) distribution 
			$\hat{\rho}^{p^\prime}$ with blue point cloud.}  
			\label{fig:Particles}
		\end{figure} 

	\item \textbf{$\bm{P}$-Dimensional Matching}: Following the discrete OT Monge formulation for the special case where both 
		distributions contain the same number of points and all weights are uniform \cite{peyre20, torregrosa2022surrogate}, 
		determine the optimal correspondence between two distributions, $\rho^p$ and $\rho^{p^\prime}$, each represented by 
		$N_{s}$ particles, by finding the optimal bijection $\phi_{s}^p$ within the set of $N_s$ \emph{permutations} 
		that minimizes the cost function 
		\[ C_{p, p^\prime}(\phi_s^p, \phi_s^{p^\prime}) = \sum_{n=1}^{N_{s}} \left\| \bm{\mu}_{\phi_s^{p}(n)}
		- \bm{\mu}_{\phi_s^{p^\prime}(n)} \right\|_2^2\] 
		(In other words, this step reorders the particle arrangements in the matrices $\bm{\mu}^p$ to reflect the optimal matching:
		each particle in the $n$-th row of one distribution is thus paired with the corresponding particle in the $n$-th row 
		of the other distributions, establishing a consistent correspondence that justifies interpolation). Interpolation is 
		then performed by partially displacing the particles along the line segments connecting matched pairs, as illustrated 
		in Figure \ref{fig:Particles}. 

		When interpolating across more than two distributions ($P > 2$), the problem extends naturally to the 
		\emph{multimarginal} OT Monge formulation \cite{peyre20}. In this case, determine the optimal assignment 
		simultaneously across all distributions, as depicted in Figure \ref{fig:hypergraph} (left). Each particle in a given 
		distribution is matched with one particle from every other distribution, minimizing the aggregate matching cost 
		illustrated in Figure \ref{fig:hypergraph} (right). The total cost function is expressed as
		\[C_{P}(\phi_s^{1}, \dots, \phi_s^{P}) = \sum_{p=1}^{P-1} \sum_{p^\prime = p+1}^{P} C_{p, p^\prime}
		(\phi_s^{p}, \phi_s^{p^\prime})\] 
		The corresponding $P$-dimensional optimal matching problem entails determining $P-1$ orderings, denoted as 
		$\phi_s^{p}$, for the $N_{s}$ particles within each distribution. Since permuting only one set of distributions 
		suffices for each pair (with the other set represented by the identity map), the problem necessitates $P-1$ 
		orderings. Therefore, the general matching problem can be formulated as 
		\begin{equation}
			\label{eq:minpb2} 
			\argmin_{\phi_s^{1}, \dots, \phi_s^{P-1}} C_{P}(\phi_s^{1}, \dots, \phi_s^{P})
		\end{equation}
		Upon obtaining the optimal solution, each particle is consistently matched across all distributions, allowing the 
		tracking of individual particles through the $P$ distributions, as shown in Figure \ref{fig:hypergraph}. It is worth 
		noting that this problem is equivalent to the \emph{Multidimensional Assignment Problem} (MAP), with an input size of 
		$N_{s}^{P}$ and a number of feasible assignments growing as $N_{s}!^{(P-1)}$, making it NP-hard for $P > 2$ 
		\cite{kammerdiner22}. A heuristic approach is therefore employed to obtain a feasible minimum.

		\begin{figure}[H]
			\centering 
			\includegraphics[width=0.8\textwidth]{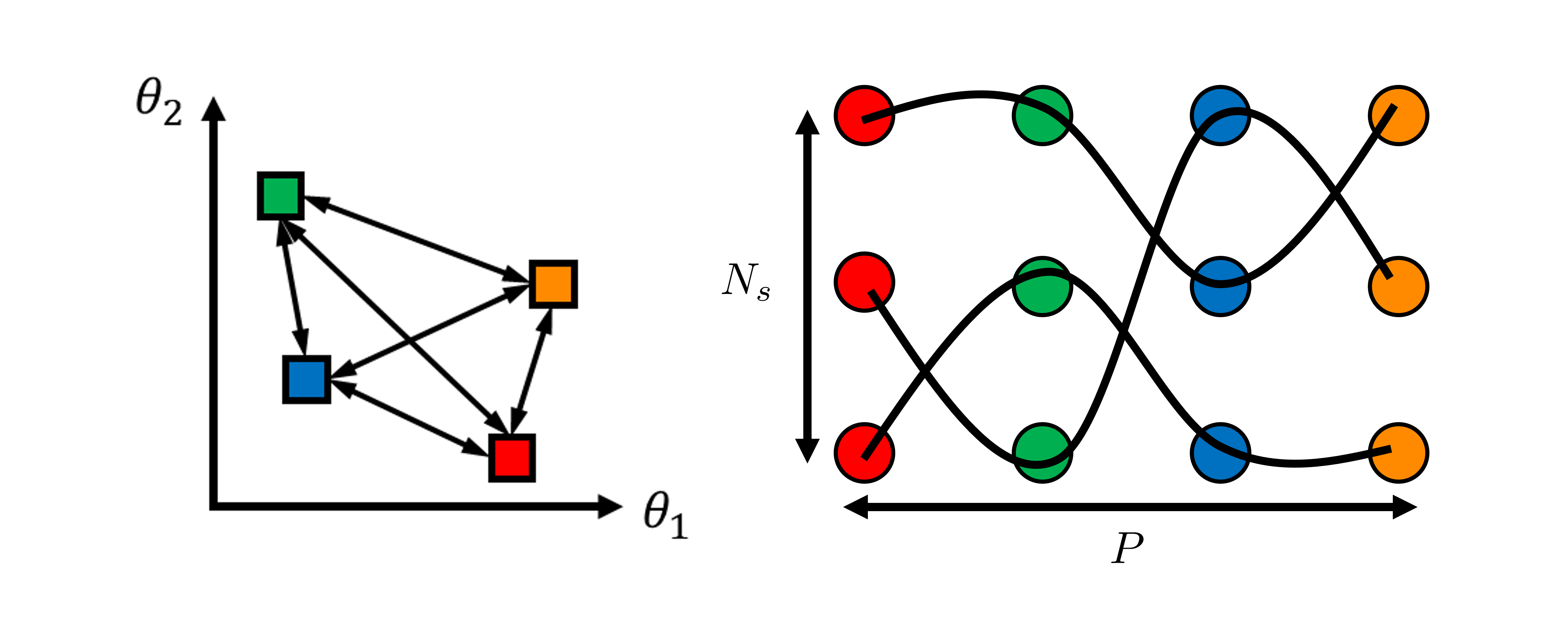} 
			\captionsetup{width=0.8\linewidth}
			\caption{Visualization of the optimal $P$-dimensional matching: four distinct distributions ($P=4$) in a 
			two-dimensional parameter domain $\mathcal{D} = (\bm{\theta} = (\theta_1, \theta_2))$, each uniquely 
			colored, with black double-headed arrows denoting the $P$-dimensional matching (left); schematic of the 
			matching process with $P=4$ and $N_{s}=3$, illustrating four colored point clouds (three points each), 
			with curved black lines representing the optimal assignments (right).} 
			\label{fig:hypergraph} 
		\end{figure} 

	\item \textbf{Regressor Training}: Train a regressor to predict the positions of $N_{s}$ particles using a training set of $P$
		distributions, i.e., $P$ matched point clouds. This enables the regressor to infer the particle positions for any 
		out-of-sample parameter vector $\bm{\theta}^\star \in \mathcal{D}$. To construct this model efficiently, non-intrusive
		model order reduction \cite{mcquarrie23} is applied to reduce the computational cost associated with the large-scale 
		particle dataset. In particular, proper orthogonal decomposition is used, computed via singular value decomposition 
		(SVD) \cite{kunisch1999}. Define a solution snapshot matrix $\bm{\Gamma} \in \mathbb{R}^{d N_{s} \times P}$, where 
		each column contains the coordinates of the $N_{s}$ particles from one of the $P$ training distributions. Apply SVD to
		$\bm{\Gamma}$ and use the standard SVD energy criterion to obtain the reduced-order basis (ROB) $\mathbf{U} \in 
		\mathbb{R}^{d N_s \times R}$, where $R < P$ (and possibly $R \ll P$). The ROB compresses the snapshot matrix as 
		\begin{equation*} 
			\bm{\Gamma} \approx \mathbf{U} \bm{\alpha}
		\end{equation*} 
		where $\bm{\alpha} \in \mathbb{R}^{R \times P}$ contains the generalized coefficients. For each mode 
		$r \in \llbracket R \rrbracket$ (i.e., $r = 1, \dots, R$), there is a corresponding vector of coefficients 
		$\bm{\alpha}_r \in \mathbb{R}^{P}$, where each element $\bm{\alpha}_{r,p}$ corresponds to one of the sampled 
		solutions $p \in \llbracket P \rrbracket$ in the parameter domain. To learn the set of entries 
		$\{\bm{\alpha}_{r,p}\}_{p=1}^{P}$ from the sampled parameter vectors $\{\bm{\theta}^p \in \mathcal{D}\}_{p=1}^{P}$, 
		a regressor is employed. The choice of regressor is not critical, provided it can capture complex dependencies in 
		high-dimensional spaces using limited data. 
\end{enumerate}

\subsection{Interpolation of Positive Scalar Field Solution Snapshots}
\label{sec:INTERP}

The primary objective of the online stage of the OT-based interpolation methodology introduced in \cite{torregrosa2022surrogate} 
is to interpolate the set of precomputed positive scalar field solution snapshots $\left\{\psi^p\right\}_{p=1}^P$ to estimate 
$\psi^\star$, the value of the scalar field $\psi$ at an out-of-sample parameter vector $\bm{\theta}^\star \in 
\mathcal{D}$. This online stage consists of the following real-time steps:
\begin{itemize}
	\item Regress the $N_s$ positions of the particles from the $P$ distributions to infer the position matrix 
		$\bm{\mu}^\star \in \mathbb{R}^{N_{s} \times 2}$, involving the displacement of all particles from the 
		precomputed solution snapshots.
	\item Reconstruct the distribution $\hat{\rho}^\star$ using the $N_s$ identical Gaussian functions with standard deviation
		$\sigma_{s}$ and the position matrix $\bm{\mu}^\star$ (see \eqref{eq:DING}).  
	\item Regress the $P$ integrals $\left\{I^p\right\}_{p=1}^P$ to infer $I^\star = \int_\Omega \psi^{\star} \, d\Omega$, 
		using the same regressor employed in the Regressor Training substep of the offline stage
		(see Section \ref{sec:OFFLINE}).
	\item Predict $\hat{\psi}^{\star} = \psi(\mathbf{X}, \bm{\theta}^\star)$ as follows
		\begin{equation*} 
			\hat{\psi}^{\star} = I^{\star} \hat{\rho}^{\star}, \quad  \text{where} \quad  \hat{\rho}^{\star}
			(\mathbf{X}) = \sum_{n=1}^{N_{s}} G_{\bm{\mu}_{n}^{\star},\sigma{s}} (\mathbf{X}) 
		\end{equation*}
\end{itemize}

From these steps, we conclude that the outlined OT-based interpolation methodology is inherently structure-preserving, as it maintains
two key physical properties of the approximated scalar fields:
\begin{itemize}
    \item \textbf{Maintenance of Form (localized features):} The methodology preserves the shape of scalar fields, particularly 
	    localized phenomena such as stress concentrations, sharp gradients, or vortices. Unlike conventional linear (Euclidean) 
		interpolation, which can introduce artificial smoothing, the OT-based approach transports ``mass'' (physical 
		quantities like energy, temperature, or pressure, interpreted as measures) in a physically meaningful way, ensuring 
		localized features are retained as they move or deform between snapshots.  
	\item \textbf{Preservation of Positive Signs:} The methodology maintains the positive (or non-negative) sign of the original 
		scalar fields. By modeling physical quantities as measures, OT theory guarantees that the interpolated field retains 
		its non-negative nature.
\end{itemize}

In summary, ``structure-preserving'' refers to the method's ability to produce predictions that retain the core physical 
characteristics of the original field -- the localized form and the sign -- through the fundamental properties of OT theory.

\subsection{Summary}

\begin{figure}[H] 
	\centering 
	\includegraphics[width=1\textwidth]{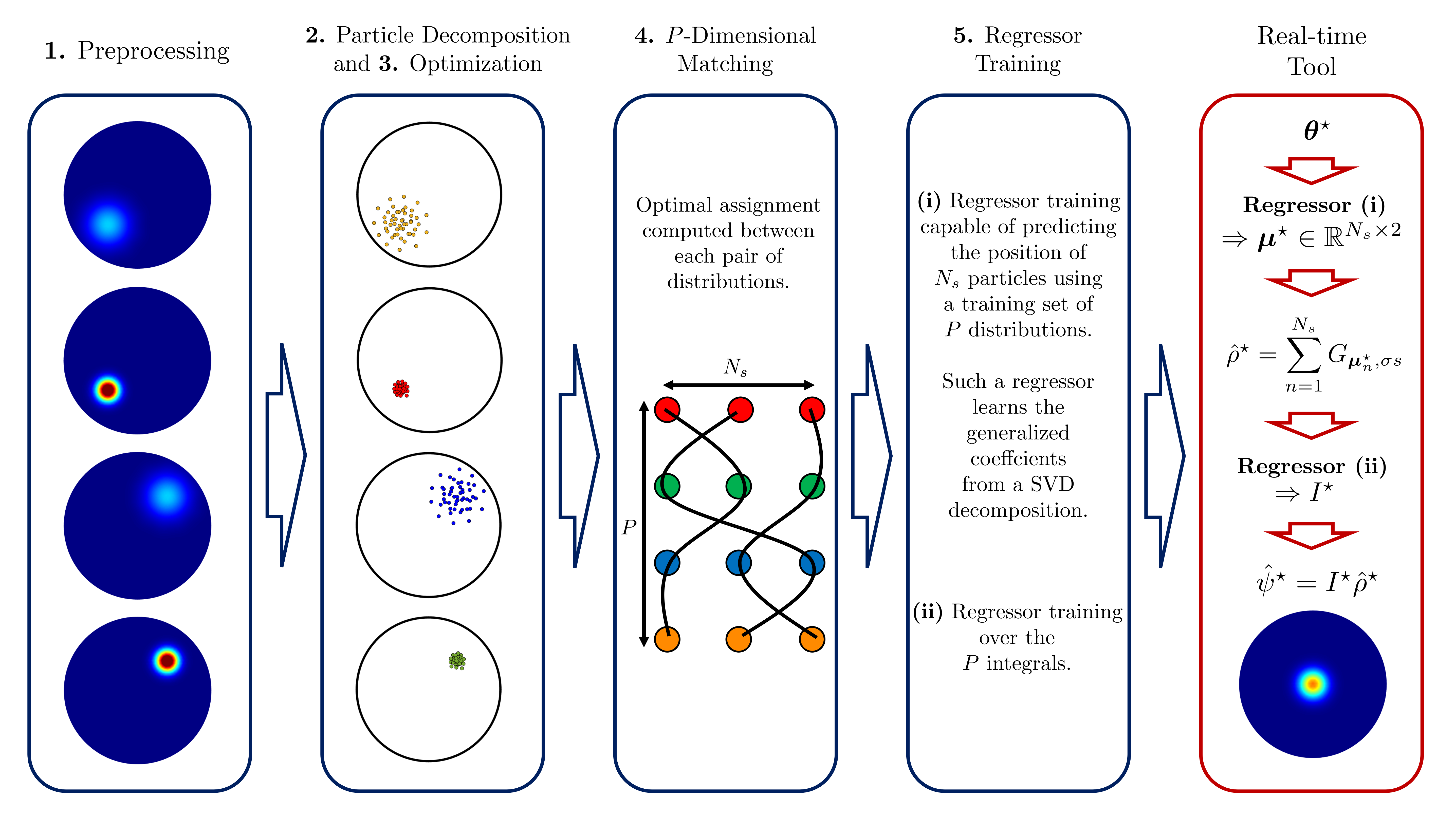} 
	\caption{Overview of the OT-based interpolation methodology: the offline stage is shown in blue, while the online stage 
	is represented in red.} 
	\label{fig:OToverview} 
\end{figure}

\begin{figure}[H] 
	\centering 
	\includegraphics[width=0.9\textwidth]{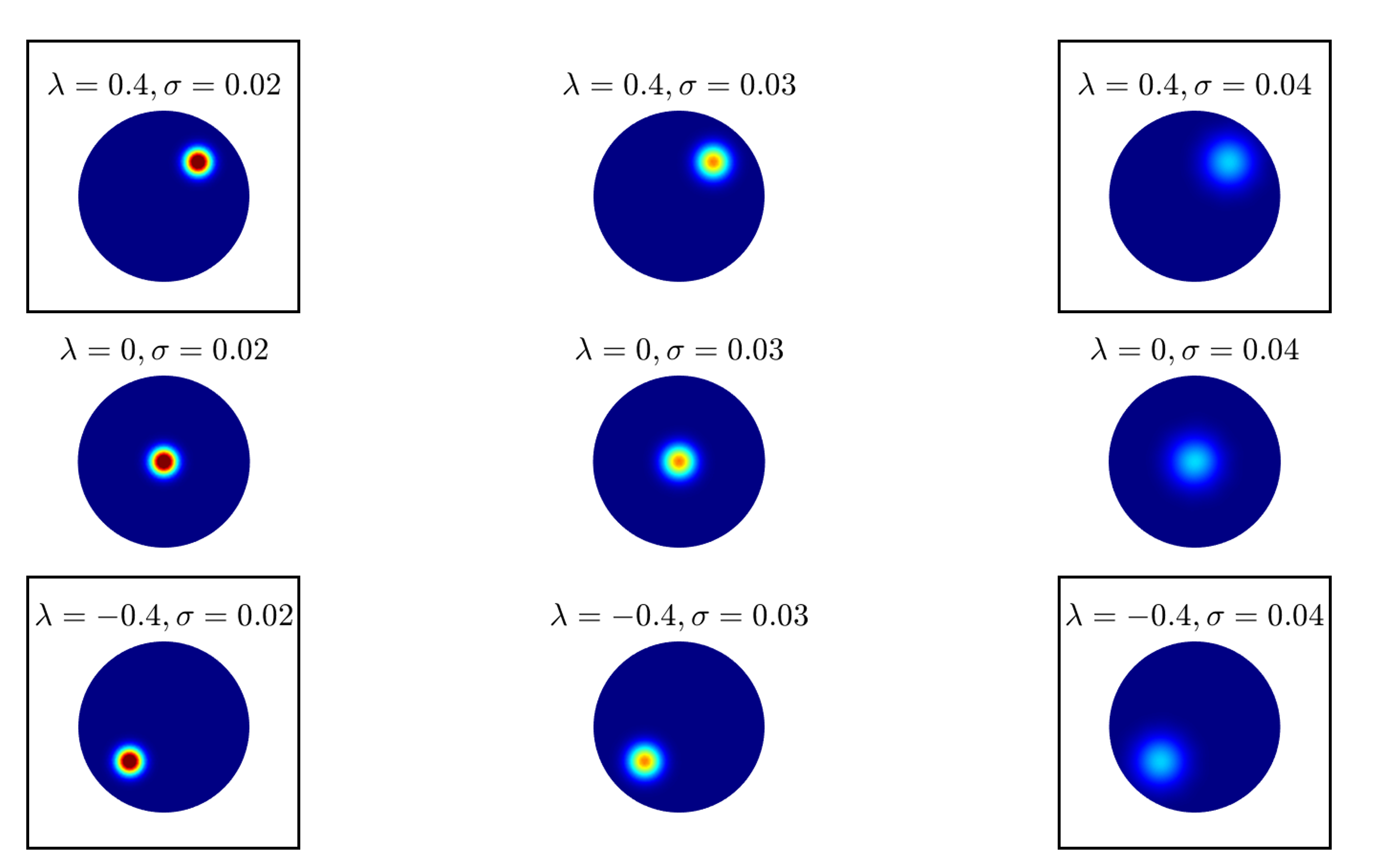} 
	\caption{Application of the OT-based interpolation methodology to solve a specific instance of problem $\mathcal{P}$ with
	$P=4$: training is conducted using the framed samples.} 
	\label{fig:ToyProblem_Sol} 
\end{figure}

Figure \ref{fig:OToverview} offers an overview of the OT-based interpolation methodology, distinguishing between the offline 
stage (in blue) and the online stage (in red). Figure \ref{fig:ToyProblem_Sol} exemplifies this approach through a specific 
instance of problem $\mathcal{P}$ with $P=4$. Together, the offline and online stages create a surrogate model 
$\hat{\psi}(\mathbf{X})$ for the positive scalar field $\psi(\mathbf{X})$, which we will refer to as the surrogate solution 
model (SSM) $\mathcal{S}$.

\section{Interpolation of Shape-Defined Positive Scalar Field Solution Snapshots Using Optimal Transport}
\label{sec:GENINTERP}

In this section, we present our methodology for geometry exploration, grounded in OT and integrated with the previous OT-based 
approach, enabling simultaneous exploration of geometries and prediction of associated positive scalar fields. This innovative 
framework allows for a comprehensive examination of geometric shapes while harnessing OT's strengths for accurate and efficient 
interpolations, especially for localized solutions.

\subsection{Exploration of Parametric Geometries}
\label{sec:EXPLORATION}

We extend the previously presented OT-based methodology, which learns parametric positive scalar solution fields, to explore 
parametric geometries -- including parameterized shapes represented by elements (or tiles) and vertices -- in one, two,
and three dimensions. Specifically, we focus on cases where the parameter domain $\mathcal{D}$ includes shape and/or other geometric 
parameters, leading to the spatial domain $\Omega$ being dependent on $\bm{\theta} \in \mathcal{D}$.

Exploring various geometries in the context of GD presents numerous advantages. For instance, in a what-if scenario, analysts can 
evaluate multiple geometries and their corresponding parametric solution snapshots, pondering how these snapshots would adapt to 
additional geometries. In design optimization, the goal is to iteratively refine the parameters that define the geometry's 
representation until the optimal configuration is identified. Achieving this requires a thorough understanding of the solution 
fields associated with each iterative design. Efficient, accurate, and practical snapshot interpolation methods become essential 
for accelerating this process.

To address these challenges, our primary approach has two complementary objectives:
\begin{itemize}
	\item \textbf{Reparameterization of Geometries}: The first objective is to reparameterize any known geometry -- such as one 
		stored in the database $\mathcal{DB}$ -- or any unknown geometry -- such as one to be inferred from the known 
		geometries -- to achieve two key goals. The first goal is to reduce the number of parameters required to represent 
		the geometry and its evolution. Initially, this representation may involve up to three parameters per vertex 
		(displacements in the $x$, $y$, and $z$ directions). The second goal is to allow geometries to have varying numbers 
		of vertices, enabling solution snapshots of different dimensions. Reducing the number of parameters improves 
		computational efficiency, while flexibility in vertex count enhances practical applicability. Both goals are achieved 
		by offline decomposition of a carefully constructed field -- based on the corresponding signed distance function (SDF)
		-- into a sum of particles (or Gaussian functions), as described in Steps 2 and 3 of Section \ref{sec:BACKPLUS}.  
		Regardless of the initial representation of the geometry (e.g., mesh, implicit surface, etc.), it is first converted 
		into an SDF. This unified representation allows geometries from heterogeneous sources to be processed consistently, 
		streamlining the reparameterization of known geometries.
	\item \textbf{Online Interpolation of Particle Coordinates}: Our second objective focuses on the online interpolation of the 
		coordinates of particles for an unknown geometry -- for example, one that is not stored in $\mathcal{DB}$. 
		This step is critical, as reparameterizing a given geometry using 
		a sum of Gaussian functions can be computationally intensive, typically necessitating the resolution of an 
		optimization problem such as \eqref{eq:minpb1}. We will refer to this process as the OT-based geometry interpolation 
		methodology. While our main approach is tailored to the design optimization scenario outlined above, it is also 
		extendable to various contexts, including the what-if scenario.
\end{itemize}

To illustrate the extended OT-based methodology for simultaneously interpolating parametric geometries and their associated 
parametric positive scalar fields, we will revisit the two-dimensional geometry depicted in Figure \ref{fig:ToyProblem} at various
stages as we progress through the presentation of the method, defining the parameters $a$ and $b$ along the way. By varying $a$ and 
$b$, we generate $K$ geometrical domains $\Omega^{k}$ for $k \in \llbracket K \rrbracket$ and store them in the database $\mathcal{DB}$, 
along with their corresponding positive scalar solution snapshots. Additionally, we redefine the problem $\mathcal{P}$, introduced in 
Section \ref{sec:OFFLINE}, as the task of interpolating two or more geometries and possibly their relevant positive scalar 
snapshots that exhibit localization.

\begin{figure}[H] 
	\centering 
	\includegraphics[width=1\textwidth]{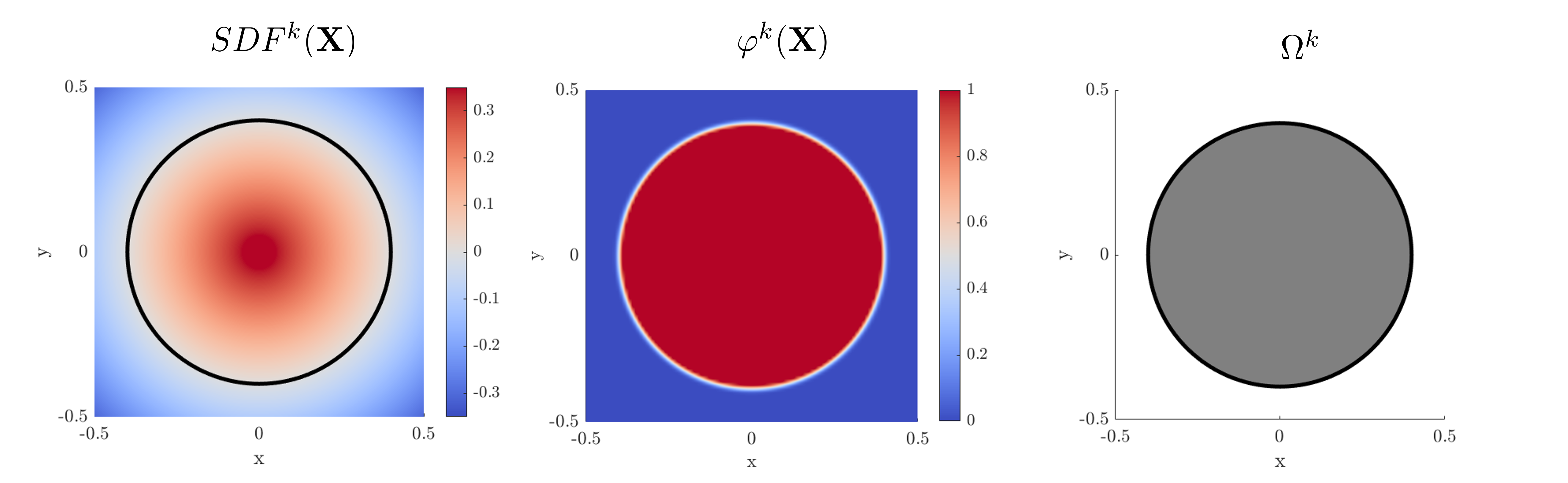} 
	\caption{Smallest signed distance to the boundary $\partial \Omega^{k}$ computed in $\Omega^\text{ref}$ (left); 
	sigmoid function $\varphi^{k}$ (middle); geometrical domain $\Omega^{k}$ (right).} 
	\label{fig:LevelSet} 
\end{figure}

Specifically, we reparameterize any geometry of interest -- referring to one of the $K$ geometries mentioned previously
-- as a function in $\mathbb{R}^{d}$ using a signed distance function (SDF). The SDF is a continuous function that outputs the 
smallest distance $d_{\partial \Omega^{k}}$ to the boundary $\partial \Omega^{k}$ for any spatial point in $\Omega^\text{ref}$, 
a reference fictitious domain that encompasses all $K$ geometries
\begin{equation*} 
	SDF^{k}(\mathbf{X}) = 
	\begin{cases} 
		d_{\partial \Omega^{k}}& ~~ \text{if} ~~ \mathbf{X} \in \Omega^{k} \\ 
		0& ~~ \text{if} ~~ \mathbf{X} \in \partial \Omega^{k} \\ 
		- d_{\partial \Omega^{k}}& ~~ \text{if} ~~ \mathbf{X} \notin \Omega^{k} \\ 
	\end{cases} 
\end{equation*}
Hence, the boundary of the geometrical domain $\Omega^{k}$ is represented by the condition $SDF^{k}(\mathbf{X}) = 0$. We then define $K$ 
sigmoid functions $\varphi^{k}$ based on these distances to characterize the $K$ geometrical domains, as follows
\begin{equation*} 
	\begin{split} 
		\varphi^{k}:  \Omega^\text{ref} & \rightarrow \left[0, 1 \right] \\ 
		\mathbf{X} & \rightarrow  \frac{1}{1+\exp\left(-SDF^{k}(\mathbf{X}) \right)}
	\end{split} 
\end{equation*}
These functions, defined on $\Omega^{\text{ref}}$, monotonically increase from 0 to 1, reaching 0.5 at the boundary, as 
illustrated in Figure \ref{fig:LevelSet}. Consequently, a point $\mathbf{X} \in \Omega^{\text{ref}}$ is considered to be inside 
$\Omega^{k}$ if $\varphi^{k}(\mathbf{X}) \geq 0.5$.

Here, we outline the offline and online stages of the OT-based geometry interpolation methodology. Some stages align with those 
involved in the construction and execution of the OT-based SSM $\mathcal{S}$ described in Sections \ref{sec:OFFLINE} and 
\ref{sec:INTERP}; we will refer back to earlier explanations in these instances.

\begin{enumerate} 
	\item \textbf{Preprocessing}: Normalize offline the functions $\varphi^{k}$ to obtain unitary integral distributions
		\begin{equation*}\label{normalization} 
			\eta^{k} = \frac{\varphi^{k}}{J^{k}},  \quad \text{where} \quad J^{k} = \int_{\Omega^{\text{ref}}} 
			\varphi^{k} \, \dd\Omega^{\text{ref}}  
		\end{equation*} 
		
			\begin{figure}[H]
			\centering
			\includegraphics[width=0.7\textwidth]{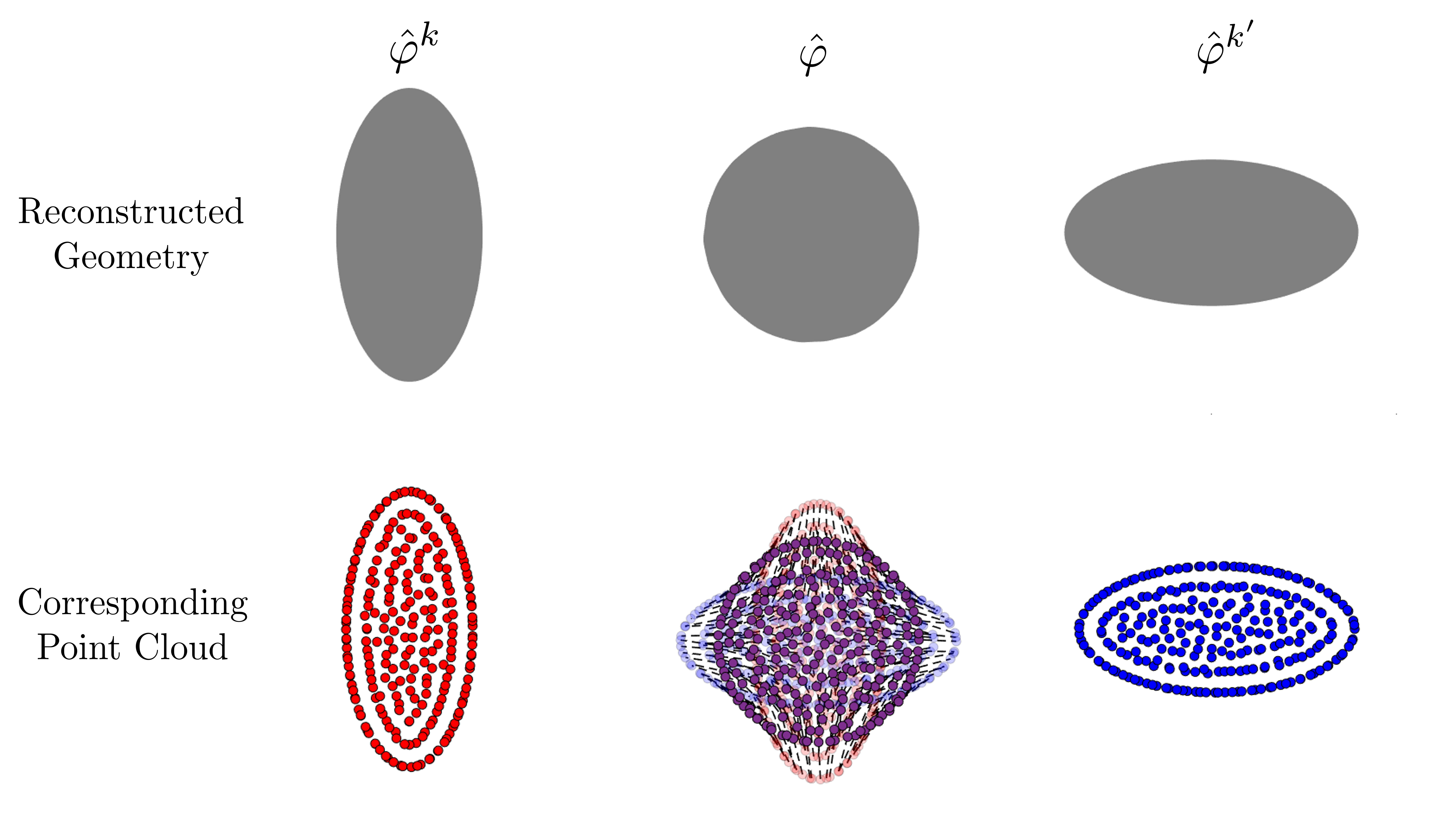}
			\caption{Reconstructed function $\hat{\varphi}^{k}$ and its corresponding point cloud in red (left); 
				partially displaced point cloud in violet along with its reconstructed function $\hat{\varphi}$ and 
				optimal matching between clouds indicated by black dotted lines (middle); reconstructed function 
				$\hat{\varphi}^{k^\prime}$ and its corresponding point cloud in blue (right).} 
			\label{fig:Particles_Geo}
		\end{figure} 
		
	\item \textbf{Particle Decomposition}: Using the procedure outlined in Steps 2 and 3 of Section \ref{sec:BACKPLUS},
		decompose, in the offline stage, each distribution associated with a given geometry into a sum of $N_{g}$ 
		two-dimensional Gaussian functions. Each Gaussian has a fixed standard deviation $\sigma_{g}$ and an integral 
		equal to $1/N_{g}$, where the subscript $g$ denotes quantities associated with geometries. Treat both $N_g$ and 
		$\sigma_g$ as hyperparameters that can be tuned during the offline stage. Then, introduce the matrix 
		$\bm{\gamma}^{k} \in \mathbb{R}^{N_{g}\times 2}$, which, for each 
		particle within the distribution $\eta^{k}$, contains the $x$ and $y$ coordinates $\gamma^k_{n,x}$ and 
		$\gamma^k_{n,y}$ of the two-dimensional vector $\bm{\gamma}^k_n$. This process parallels that of $\bm{\mu}^{p}$, 
		as illustrated in Figure \ref{fig:Particles_Geo}. The reconstructed distribution $\hat{\eta}^{k}$ is given by 
		\begin{equation} 
			\label{eq:reparag}
			\hat{\eta}^{k}(\mathbf{X}) =  \sum_{n=1}^{N_{g}} G_{\bm{\gamma}_{n,\sigma_{g}}^{k}}(\mathbf{X}),   \quad 
			\text{where} \quad G_{\bm{\gamma}_{n,\sigma_{g}}^{k}}(\mathbf{X}) = \frac{1}{N_{g}\sigma_{g}^{2}2\pi} 
			\exp\left(-\frac{\left(\mathbf{X}-\bm{\gamma}_{n}^{k}\right)^{2}}{2\sigma_{g}^{2}}\right) 
		\end{equation} 

	\item \textbf{$\bm{K}$-Dimensional Matching}: Determine offline the optimal matching between two distributions, $\eta^k$ and 
		$\eta^{k^\prime}$, each approximated by a decomposition into $N_s$ particles $\bm{\gamma}^{k}$, where $k \in 
		\llbracket K \rrbracket$. The corresponding optimization problem mirrors that addressed in Step 4 of Section 
		\ref{sec:OFFLINE}, where each particle from one distribution is matched with exactly one particle from the other 
		distribution, minimizing the defined cost $C_{K}$ as shown in \eqref{eq:minpb2}. This approach 
		facilitates robust and accurate interpolation between the distributions through the partial displacement of all 
		particles, as illustrated in Figure \ref{fig:Particles_Geo} for two distributions.

	\item \textbf{Interpolation}: Perform online interpolation of geometries stored in $\mathcal{DB}$ using the 
		reparameterization in \eqref{eq:reparag} to directly obtain the reparameterization of a new geometry of interest.  
		For this purpose, adopt a Wasserstein barycentric approach \cite{agueh11}, which provides a framework for computing
		the barycenter (or ``average'') of a set of distributions using the Wasserstein distance metric.  

		Specifically, given a set of input distributions $\eta^{k}$, for $k \in \llbracket K \rrbracket$, defined on 
		$\Omega^{\text{ref}}$, compute the Wasserstein barycenter for a family of weights $\omega^{k}$ that define the 
		new geometry of interest (see {\it REMARK 1}), where $\sum_{k=1}^{K} \omega^k = 1$.  

		Using the special case of the discrete multimarginal OT Monge formulation \cite{anderes16} and the previously computed
		optimal matching of the point clouds $\bm{\gamma}^{k}$, $k \in \llbracket K \rrbracket$, the Wasserstein barycenter 
		problem can be expressed as 
		\begin{equation*} 
			\argmin_{\bm{\gamma}} \sum_{k=1}^{K} \omega^{k} C_{\bm{\gamma}, \bm{\gamma}^{k}}, 
			\quad \text{where} \quad C_{\bm{\gamma}, \bm{\gamma}^{k}} = \sum_{i=1}^{N_{g}} \lVert \bm{\gamma}_{\phi_g(i)} 
			- \bm{\gamma}^{k}_{\phi_g^k(i)} \rVert_{2}^{2} 
		\end{equation*} 
		Here, the optimal bijection $\phi_g^k$ for geometries plays a role analogous to that of the optimal bijection 
		$\phi_{s}^{p}$ for solution snapshots. The solution of this minimization problem is given by the weighted 
		linear average \cite{peyre20} 
		\begin{equation} 
			\label{eq:YAP} 
			\bm{\gamma}^{\star} = \bm{\gamma}^{\mathrm{wbc}} \equiv \sum_{k=1}^{K} \omega^{k} \bm{\gamma}^{k}_{\phi_g^k}
		\end{equation} 
		The corresponding distribution $\hat{\eta}^{\star}$ can then be reconstructed by summing all Gaussian functions 
		\begin{equation*} 
			\hat{\eta}^{\star}(\mathbf{X}) = \sum_{n=1}^{N_{g}} G_{\bm{\gamma}^{\star}_{n}, \sigma_{g}}(\mathbf{X})
		\end{equation*} 

		Our focus is on geometrical reconstruction, specifically on the functions $\varphi^{k}$ that define the geometrical 
		domains $\Omega^{k}$ via their 0.5 iso-contours. By leveraging the bounded nature of these functions (between 0 and 1),
		we scale $\hat{\eta}^{\star}$ by its maximum value to obtain the reconstructed level-set function 
		\begin{equation*} 
			\hat{\varphi}^{\star}(\mathbf{X}) = \frac{\hat{\eta}^{\star}(\mathbf{X})}{\max \hat{\eta}^{\star}(\mathbf{X})}
		\end{equation*}	
		This approach eliminates the need to train a regressor for $J^{\star}$ -- the integral at the out-of-sample
		parameter point of interest -- as was previously required for constructing the SSM $\mathcal{S}$ in Sections 
		\ref{sec:OFFLINE} and \ref{sec:INTERP}.
\end{enumerate} 

{\it REMARK 1}. Given a database $\mathcal{DB}$ of sampled geometries and their reparameterizations using sums of Gaussian functions, 
the barycentric description in \eqref{eq:YAP} provides yet another representation of a parametric geometry. This description is 
applicable in various scenarios relevant to GD, including design optimization. In this context, it defines a space of admissible 
geometries that can be explored to identify optimal configurations. However, representing a specific out-of-sample geometry 
in the form provided in \eqref{eq:YAP} may necessitate an additional layer of approximation. In all cases, the OT barycenter strategy 
employed here benefits from an original parametric representation of the sampled geometries, which facilitates the application of the 
intermediate representation in \eqref{eq:YAP} -- with the final representation being a sum of Gaussian functions. This principle holds 
true in all machine learning processes, where a thoughtful selection and representation of the database content enhances the 
effectiveness of surrogate modeling.

Together, the offline and online stages described above create a surrogate model $\hat{\varphi}(\mathbf{X})$ for an out-of-sample
geometry of interest $\varphi(\mathbf{X})$, which we will denote as the surrogate geometric model (SGM) $\mathcal{G}$.

\begin{figure}[H] 
	\centering 
	\includegraphics[width=0.9\textwidth]{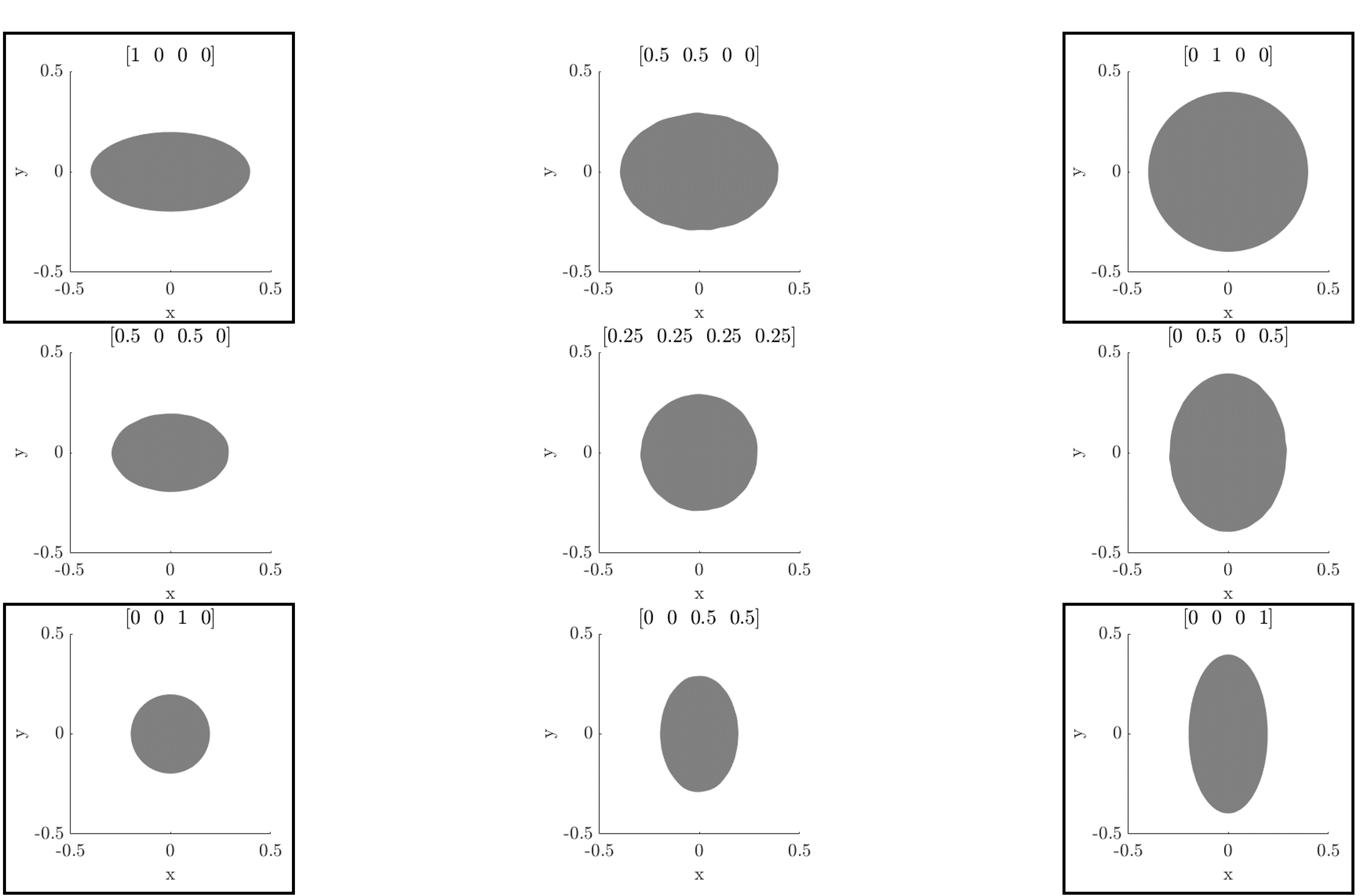} 
	\caption{Application of the OT-based geometry interpolation methodology to a specific instance of problem $\mathcal{P}$ with 
	$K=4$ samples: the four sample geometries are framed, the out-of-sample (interpolated) geometries are unframed, and the 
	corresponding weights $\omega^{k}$, $k \in \llbracket 4 \rrbracket$, are shown as $[\omega^{1} \quad \omega^{2} \quad 
	\omega^{3} \quad \omega^{4}]$.}
	\label{fig:OT_Geo} 
\end{figure}

The proposed OT-based geometry interpolation methodology is illustrated in Figure \ref{fig:OT_Geo} for a specific instance of the
redefined problem $\mathcal{P}$ with $K=4$.

\subsection{Interpolation of Parametric Geometries and Positive Scalar Field Solution Snapshots}

Thus far, we have presented an OT-based interpolation methodology for two distinct purposes:
\begin{enumerate} 
	\item To construct, in Section \ref{sec:BACKPLUS}, a real-time parametric SSM $\mathcal{S}$ for a given geometry
		that infers a positive scalar field at an out-of-sample parameter vector $\bm{\theta}^\star$ within the 
		parameter domain $\mathcal{D}$. This inference leverages the knowledge of counterpart fields precomputed at 
		sampled parameter vectors and stored in a database $\mathcal{DB}$.
	\item To construct, in Section \ref{sec:EXPLORATION}, a real-time SGM $\mathcal{G}$ for interpolating geometrical
		domains -- more specifically, their reparameterizations using sums of Gaussian functions.
\end{enumerate}
Here, we combine both strategies to develop a real-time OT-based approach that offers a parametric solution for any explored 
geometry. This approach incorporates four main hyperparameters: $N_{s}$ and $N_{g}$, which represent the numbers of particles 
for the decomposition of solutions and geometries, respectively, along with $\sigma_{s}$ and $\sigma_{g}$ -- the standard
deviations for these decompositions. These hyperparameters can be tuned during an offline stage.

Specifically, we start from $K$ sampled geometrical domains $\Omega^{k}$ for $k \in \llbracket K \rrbracket$, each with a 
corresponding set of precomputed positive scalar field solution snapshots, all stored in $\mathcal{DB}$. For each domain, we train
offline the reparameterizations in \eqref{eq:reparag} and \eqref{eq:DING} to construct an SGM $\mathcal{G}$ and its associated
SSM $\mathcal{S}^{k}$, which are then added to the content of $\mathcal{DB}$. This training may involve in each domain a variable 
number of solution snapshots and has a two-fold primary objective: to enable the interpolation of positive scalar field solution 
snapshots for a sampled geometry with out-of-sample non-geometric parameters, and to facilitate the 
simultaneous interpolation of geometries and corresponding positive scalar field solution snapshots for out-of-sample geometric 
and non-geometric parameters, utilizing information stored in $\mathcal{DB}$.

\begin{figure}[H] 
	\centering 
	\includegraphics[width=0.8\textwidth]{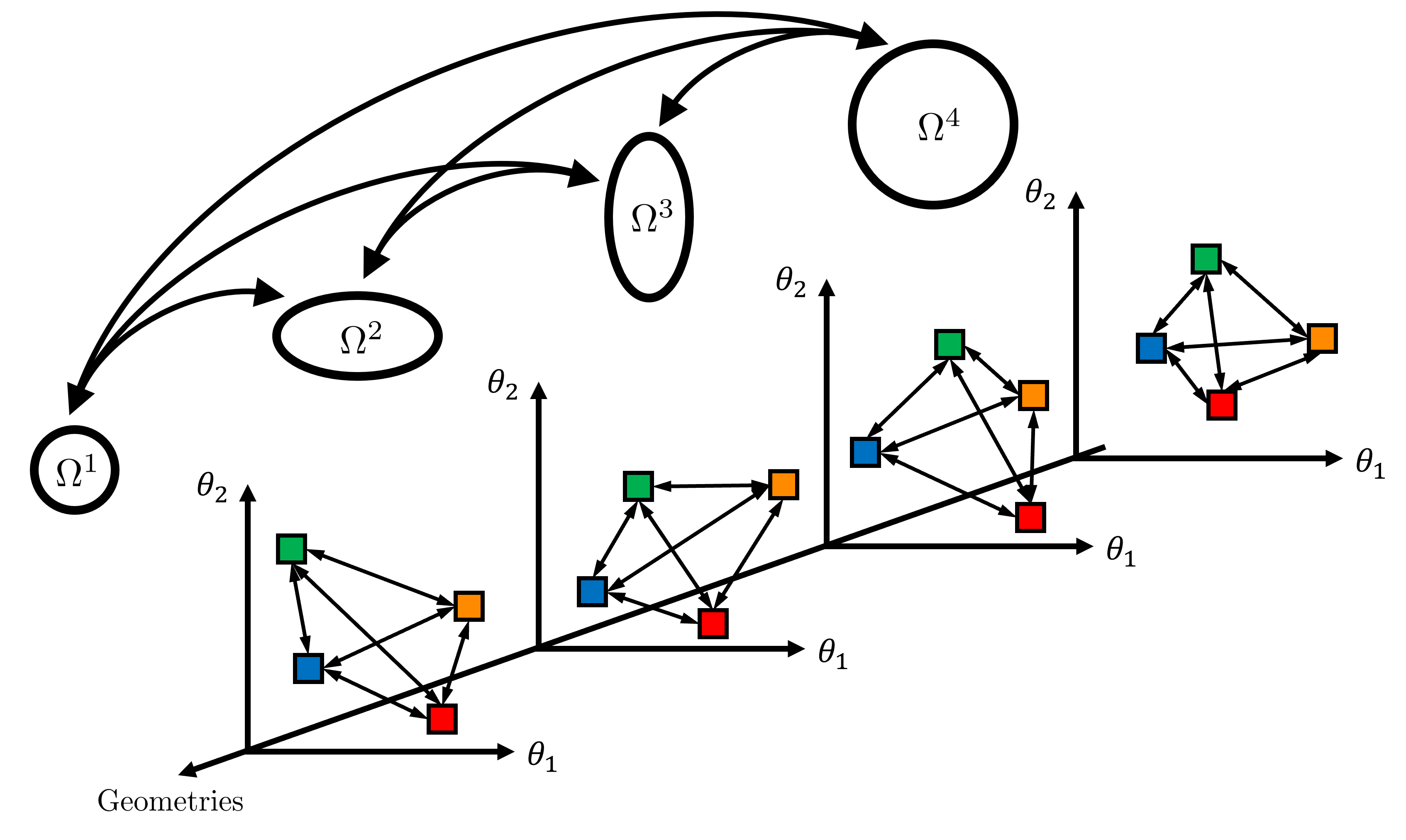} 
	\caption{Four distributions $(P=4)$ in a two-dimensional parameter domain $\mathcal{D}$ $(\bm{\theta} = (\theta_1, \theta_2))$ 
	corresponding to four different geometrical domains $\Omega^{k}$ ($K=4$), where each distribution is color-coded based on its 
	non-geometric parameter set; black double-headed straight arrows indicate matching among distributions for a given
	geometry, while black double-headed curved arrows signify matching across geometries.}
	\label{fig:PMatching_Geo} 
\end{figure}

To achieve this, the optimal matching problem during the offline training stage must consider all distributions from all 
domains, as illustrated in Figure \ref{fig:PMatching_Geo}. Consequently, the $P$-dimensional matching problem outlined in
Section \ref{sec:OFFLINE} expands to a $(K \times P)$-dimensional one. The formulation of the resulting optimal assignment 
problem \eqref{eq:minpb2} remains unchanged.

\begin{figure}[H] 
	\centering 
	\includegraphics[width=1\textwidth]{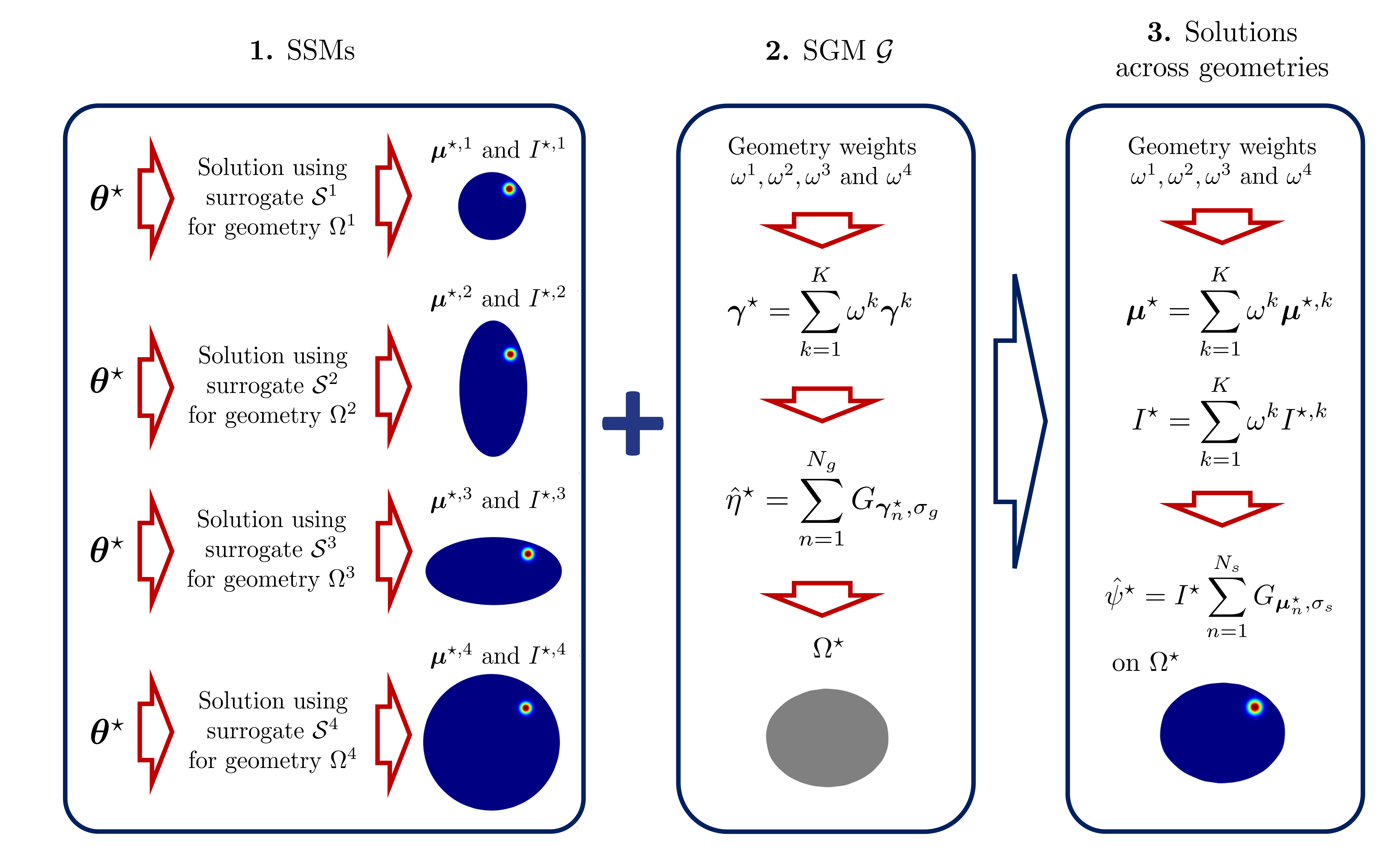} 
	\caption{Overview of the online stage of the OT-based interpolating methodology integrating an SGM and multiple SSMs 
	($K=4$).}
	\label{fig:OToverview_geo} 
\end{figure}

The online stage of the OT-based interpolation methodology combining two distinct types of surrogate models -- SGM and SSM -- is 
illustrated in Figure \ref{fig:OToverview_geo} and comprises the following real-time steps:
\begin{enumerate} 
	\item First, for a queried parameter vector $\bm{\theta}^{\star}$ with out-of-sample non-geometric parameters, apply 
		each SSM $\mathcal{S}^{k}$, $k \in \llbracket K \rrbracket$ to infer a solution $\bm{\mu}^{\star,k}$ and an 
		integral $I^{\star,k}$ in the corresponding geometrical domain $\Omega^{k}$, leading to the interpolated solutions 
		\begin{equation*} 
			\hat{\psi}^{{\star},k} = I^{{\star},k} \hat{\rho}^{{\star},k}, \quad \text{where} \quad 
			\hat{\rho}^{{\star},k}(\mathbf{X}) = \sum_{n=1}^{N_{s}} G_{\bm{\mu}^{\star,k}_{n},\sigma_{s}}
			(\mathbf{X}) 
		\end{equation*} \item Next, interpolate the reparameterization of an out-of-sample geometrical domain $\Omega^{\star}$ defined by a family
		of weights $\omega^{k}$, $k \in \llbracket K \rrbracket$, using the barycentric approach 
		\begin{equation*} 
			\bm{\gamma}^{\star} = \sum_{k=1}^{K} \omega^{k} \bm{\gamma}^{k}_{\phi_g^k} \quad \text{and} \quad \hat{\eta}^{\star} 
			= \sum_{n=1}^{N_{g}} G_{\bm{\gamma}_{n,\sigma_{g}}^{\star}}
		\end{equation*} 
	\item Subsequently, interpolate the particles and solution integral for $\bm{\theta}^{\star}$ across the geometries. Note
		that the parametric positive scalar field solution snapshots inferred from the OT-based SSMs $\mathcal{S}^k$
		for $k \in \llbracket K \rrbracket$ correspond to the same non-geometric parameters (dimensions of $\mathcal{D}$)
		across different domains. Since these snapshots are reconstructed using Gaussian functions, apply a barycentric 
		approach with the same weights as those used for the sampled geometries. While the justification for this choice 
		is most evident for small to moderate geometrical perturbations -- where the behavior of the solution snapshots is
		closely correlated with the behavior of the geometry -- this approach is applied in this work irrespective of the
		magnitude of the perturbations. Consequently, it follows that
		\begin{equation*} 
			\bm{\mu}^{\star} = \sum_{k=1}^{K} \omega^{k} \bm{\mu}^{\star,k}_{\phi_s^{p,k}} \quad \text{and} 
			\quad I^{\star} = \sum_{k=1}^{K} \omega^{k} I^{\star,k} 
		\end{equation*} 
		where $\phi_s^{p,k}$ represents an extension of the optimal bijection $\phi_s^{p}$ defined for a single geometry, now applied to 
		the ensemble of solution snapshots across all domains $\Omega^k$. 
		Finally, reconstruct the solution $\hat{\psi}^{\star}$ associated with the new geometrical domain $\Omega^{\star}$ as follows
		\begin{equation*} 
			\hat{\psi}^{\star} = I^{\star} \sum_{n=1}^{N_{s}} G_{\bm{\mu}^{\star}_{n},\sigma_{s}}  
		\end{equation*} 
\end{enumerate}

Consider the general case of a parameter vector $\bm{\theta} = (\bm{\theta}_\mathrm{nge}, \bm{\theta}_\mathrm{ge}) \in 
\mathcal{D}$, where $\bm{\theta}_\mathrm{nge}$ and $\bm{\theta}_\mathrm{ge}$ denote the subvectors of non-geometric and 
geometric parameters, respectively. Given a database $\mathcal{DB}$ containing $K$ geometrical domains $\Omega^k$ for 
$k \in \llbracket K \rrbracket$, each associated with $P^k$ positive scalar field solution snapshots, and a queried parameter 
vector $\bm{\theta}^\star = (\bm{\theta}_\mathrm{nge}^\star, \bm{\theta}_\mathrm{ge}^\star) \in \mathcal{D}$, the execution of 
this OT-based interpolation methodology, which integrates an SGM and multiple SSMs, can be implemented as follows: 
\begin{itemize}
	\item For each geometrical domain $\Omega^k$ (where $k \in \llbracket K \rrbracket$), apply the SSM $\mathcal{S}^k$ to infer the 
		positive scalar field solution snapshot at the out-of-sample parameter subvector 
		$\bm{\theta}_\mathrm{nge}$.
	\item Utilize the SGM $\mathcal{G}$, constructed using the geometric content of $\mathcal{DB}$, to interpolate the 
		geometry at the out-of-sample parameter subvector $\bm{\theta}_\mathrm{ge}$ -- specifically, 
		its reparameterization using sums of Gaussian functions.  
	\item Interpolate the desired postive scalar field solution snapshot across the reparameterized geometries utilizing the 
		set of SSMs $\left \{\mathcal{S}^k \right \}_{k=1}^K$ along with the SGM $\mathcal{G}$.
\end{itemize}

\begin{figure}[H] 
	\centering 
	\includegraphics[width=0.9\textwidth]{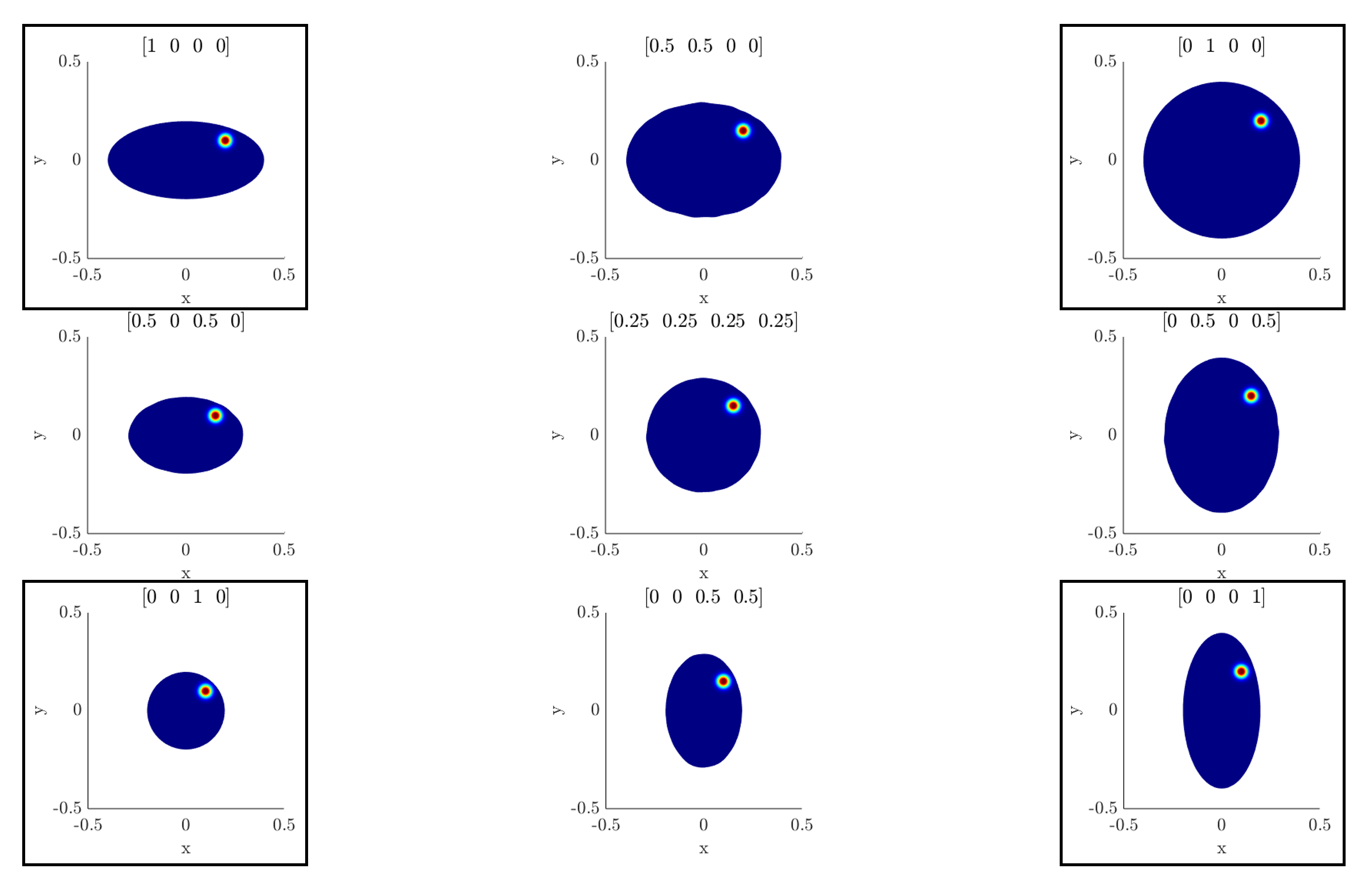} 
	\caption{Application of the OT-based interpolation methodology, which integrates an SGM and multiple SSMs, to a specific 
	instance of problem $\mathcal{P}$ with $K=4$ samples: the parametric solution is inferred within the four framed domains 
	where the surrogates were constructed, and then applied to interpolate across multiple unframed, out-of-sample geometries.}
	\label{fig:OT_Sol_Geo} 
\end{figure}

This OT-based interpolation methodology is illustrated in Figure \ref{fig:OT_Sol_Geo} for a specific instance of the redefined 
problem $\mathcal{P}$ with $K=4$.
	
\section{Extension to Parametric Arbitrarily Signed and Vector Field Solution Snapshots}
\label{sec:EXT}

In many applications, the fields of interest are often vector-valued. In structural mechanics, these include displacement, velocity, and acceleration fields, 
while in fluid mechanics, examples encompass velocity, vorticity, and force fields. Beyond being vector-valued, these fields can also possess arbitrary signs. 
Consequently, to apply the OT-based methodology developed in this paper -- designed to offer a parametric solution for any explored geometry -- certain extensions 
are necessary. 

One straightforward method for addressing the case of a vector field $\mathbf{v} \in (\mathbb{R}^+)^{\mathbbm{d}}$ is to apply the OT-based interpolation technique 
presented in this paper to each of its $\mathbbm{d}$ components. While this approach is simple, it carries the drawback of increasing the computational cost of the offline 
stage of the proposed methodology, as it necessitates solving the matching problem \eqref{eq:minpb2} $\mathbbm{d}$ times. However, considering that $\mathbbm{d}$ typically ranges 
from 2 to 6, this complication -- affecting only the training stage -- is not a significant barrier. Additionally, this limitation is commonly observed in various data-driven 
surrogate modeling techniques, including modern response surfaces and Gaussian process regressions that utilize optimal acquisition functions. 

An equally straightforward approach for handling an arbitrarily signed vector field $\mathbf{v} \in \mathbb{R}^{\mathbbm{d}}$ -- specifically for any of its components $v_i \in 
\mathbb{R}$, where $i \in \llbracket \mathbbm{d} \rrbracket$ -- is to first split $v_i$ as follows
\begin{equation}
	\label{eq:DECOMP}
	v_i = v_i^+ + v_i^- = v_i^+ - (-v_i^-), \quad \text{where} \, v_i^+ = \max(v_i, 0) \ge 0, \, v_i^- = \min(v_i, 0) \le 0, \quad \text{and} \, -v_i^- = \max(-v_i, 0) \ge 0 
\end{equation}
Then, the proposed OT-based interpolation methodology can be applied separately to $v_i^+$ and $-v_i^-$, allowing the two resulting approximations to be subtracted.

By design, the extension to vector fields of the OT-based interpolation methodology outlined above is as robust as its counterpart for scalar fields. For this reason, only
the performance of methodology for scalar fields is assessed in Section \ref{sec:APP}, including in the context of geometry exploration.

On the other hand, the extension to arbitrarily signed fields, whether scalar or vector, raises a concern regarding the use of superposition in the context of OT-based 
interpolation, which we wish to acknowledge. OT-based interpolation minimizes transportation costs to morph one distribution into another, a process that inherently involves 
nonlinearity. This nonlinearity arises from the geometric and structural properties of the distributions, which can significantly influence the resulting interpolated values.
By splitting $v_i$ into its positive and negative components, as shown in \eqref{eq:DECOMP}, and applying OT-based interpolation, each component is treated separately. Although 
these components can be processed independently, their interactions -- how they combine -- may not adhere to the linear principles applicable to traditional interpolation methods.
While subtracting the interpolated surrogates of $v_i^+$ and $-v_i^-$ can produce a function that effectively captures the essential characteristics of $v_i$, this should not be 
misinterpreted as strictly following the superposition principle of linear interpolation. The transport dynamics of each component may exhibit complex behaviors that are not linear.
Thus, although the suggested approach for extending our OT-based interpolation methodology to arbitrarily signed fields involves a difference combination, it is essential to recognize 
that the resulting interpolation does not signify linear behavior in the underlying process. Instead, it reflects an adaptation of the methodology originally designed for positive 
fields to accommodate arbitrary signs effectively. This approach requires validation within the framework of OT principles, which will be explored in further research. This ongoing 
research will also explore alternative methods for combining the interpolations of $v_i^+$ and $-v_i^-$ beyond simple subtraction. Additionally, it will investigate the possibility of 
avoiding splitting altogether by shifting $v_i$ to ensure that the resultant field remains positive. Consequently, this extension is not assessed or demonstrated in Section 
\ref{sec:APP}.

\section{Application Examples}
\label{sec:APP}

In this section, we illustrate, demonstrate, and evaluate the unified, structure-preserving computational methodology introduced 
herein. This methodology leverages OT to explore parametric geometrical domains and interpolate positively-valued scalar fields. To 
ensure reproducibility and clarity, we employ a simple yet representative set of problems. Across all cases, the following 
computational settings are consistently applied:
\begin{itemize} 
	\item The optimization problem \eqref{eq:minpb1} is solved using a gradient-based algorithm.  
	\item A genetic algorithm is employed for solving the optimization problem \eqref{eq:minpb2}.  
	\item The Regressor Training substep (detailed in Section \ref{sec:OFFLINE} of the offline stage within the OT-based 
		interpolation methodology described in Section \ref{sec:BACKPLUS}) utilizes an sPGD regressor \cite{ibanez18}.
	\item All computations are performed using double precision arithmetic on a single core of a 12th Gen Intel(R) Core(TM) i7-12650H 2.30 GHz CPU.  
\end{itemize}

Additionally, the following convergence criteria, where $i$ designates the $i$-th iteration, are adopted: 
\begin{itemize} 
	\item Particle Decomposition 
		\begin{equation*} 
			\frac{1}{2} \left\| \rho^p - \hat{\rho}_{i}^{p} \right\|_{2}^{2} -  \frac{1}{2} \left\| \rho^p - \hat{\rho}_{i+1}^{p} \right\|_{2}^{2} < 10^{-4}
		\end{equation*} 
	\item $P$-Dimensional Matching 
		\begin{equation*} 
			C_{P}^{i} - C_{P}^{i+1} < 10^2
		\end{equation*} 
		Although a threshold of $10^{2}$ may appear large in absolute terms, the initial cost is orders of magnitude higher. 
		Empirical testing indicates that lower minima are not computationally achievable. Given the heuristic nature and high 
		complexity of the genetic algorithm used to solve this optimization problem, this threshold provides a pragmatic 
		balance between solution quality and computational efficiency.
	\item SSM Training (where $f$ represents the regressor and $x$ the training inputs)
		\begin{equation*} 
			\|\hat{f}_{i}(x) - \hat{f}_{i+1}(x)\|_{2} < 10^{-2}
		\end{equation*} 
\end{itemize} 

\subsection{Parametric Heat Transfer Problem and Data Generation}
	
\begin{figure}[H] 
	\centering \includegraphics[width=0.4\textwidth]{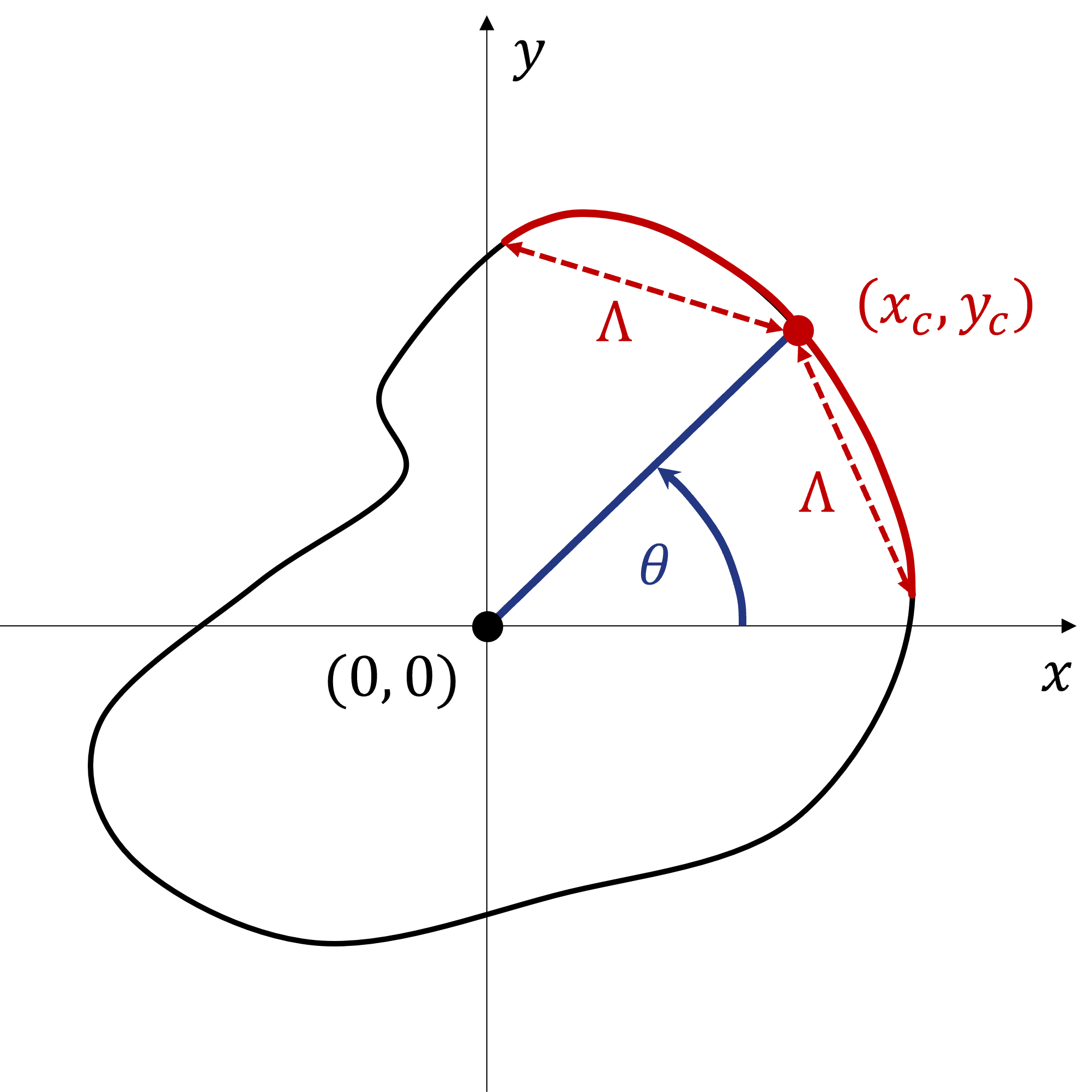} 
	\caption{Geometrical domain and parametric support of the temperature boundary condition.}
	\label{fig:Pb_results} 
\end{figure}

This study addresses a parametric transient heat transfer problem within two-dimensional, randomly generated geometrical domains 
$\Omega^k$ characterized by isotropic thermal conductivity. The problem's parametric nature stems from the temperature boundary 
condition, which is applied parametrically on a portion of the geometrical domain's boundary, as depicted in Figure 
\ref{fig:Pb_results}.

A Dirichlet boundary condition is applied, where the boundary temperature $T_{\partial \Omega^{k}}$ decreases from a specified point $(x_{c},y_{c})$ on the boundary 
$\partial \Omega^{k}$, reaching zero at a distance $\Lambda$ (see Figure \ref{fig:Pb_results}). The formulation of the problem is as follows
\begin{equation} 
	\begin{array}{r c l l} 
		\displaystyle{\kappa \frac{\partial T}{\partial t}} &=& \displaystyle{\frac{\partial^{2} T}{\partial x^{2}} + \frac{\partial^{2} T}{\partial y^{2}}} & \text{in  } 
		\Omega^{k} \times [t_{0},t_{f}]\\ [1em]
		T(x,y,t) & = &  \max\left(0, \Lambda-\| (x_{c},y_{c}) - (x,y)\|_{2}^{2}\right )/\Lambda &  \text{on }\partial \Omega^{k} \times [t_{0},t_{f}]\\ [1em]
		T(x,y,t_{0}) & = & 0 & \text{in  } \Omega^{k} 
	\end{array} 
\label{eq:heat}
\end{equation}
Here, $\kappa=0.015$ represents the diffusivity, while the initial and final times are set to $t_{0}=0$ and $t_{f}=1$, respectively. The location $(x_{c},y_{c})$ is parameterized by 
the angle $\theta$, which, along with the distance $\Lambda$, defines the problem's two-dimensional parameter domain $\mathcal{D}$. The focus is on predicting the temperature field 
$T$ within any domain at the final time $t_{f}=1$.

\begin{figure}[H]    
	\centering    
	\includegraphics[width=1.00\textwidth]{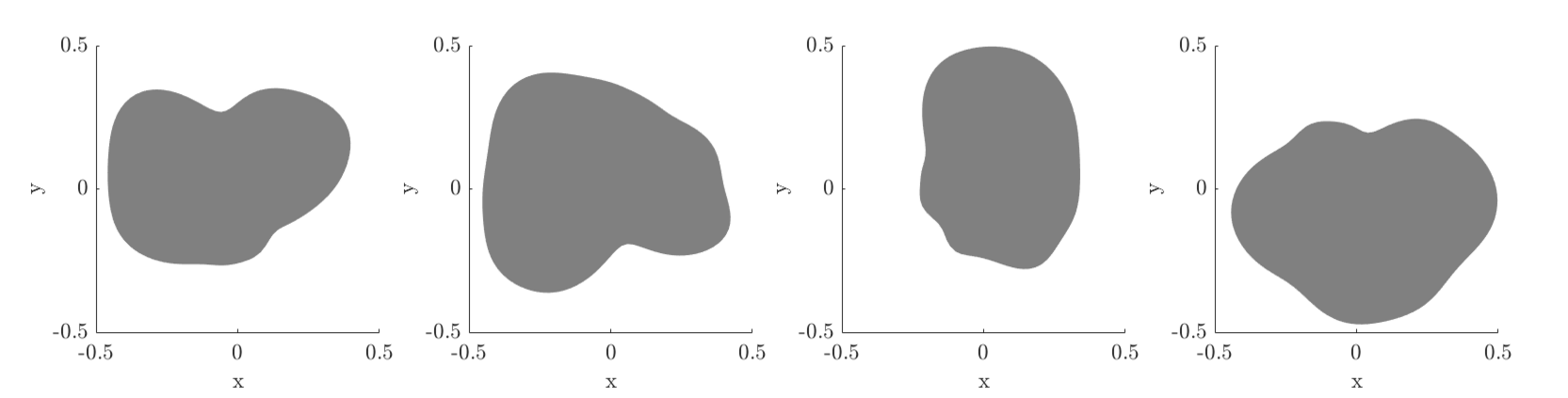}    
	\caption{Shapes of the four generated geometries -- from left to right: $\Omega^{1}$; $\Omega^{2}$; $\Omega^{3}$; and $\Omega^{4}$.}    
	\label{fig:Geometries} 
\end{figure}

We randomly generate $K=4$ geometrical domains $\Omega^{k}$, $k \in \llbracket 4 \rrbracket$, centered around the point $(0, 0)$, as illustrated in Figure \ref{fig:Geometries}. These 
geometries are crafted to demonstrate our approach's capability to handle any geometry, regardless of its creation process. 

\begin{figure}[H]
	\centering
	\includegraphics[width=1.00\textwidth]{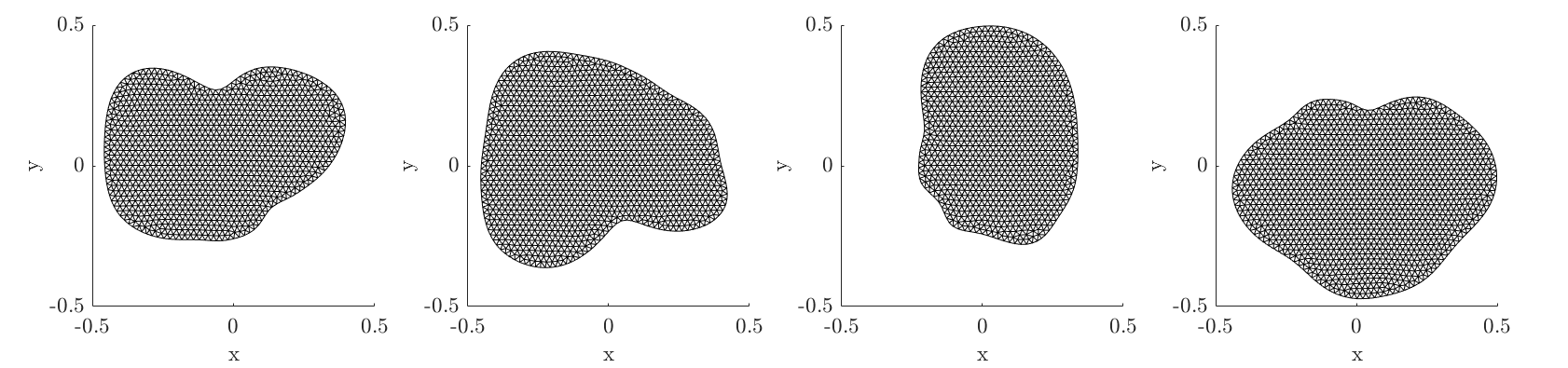}
	\caption{Mesh configurations for the four generated geometries -- from left to right: $\Omega^{1}$; $\Omega^{2}$; $\Omega^{3}$; and $\Omega^{4}$.}
	\label{fig:Geometries_mesh}
\end{figure}

\begin{figure}[H] 
	\centering 
	\includegraphics[width=1\textwidth]{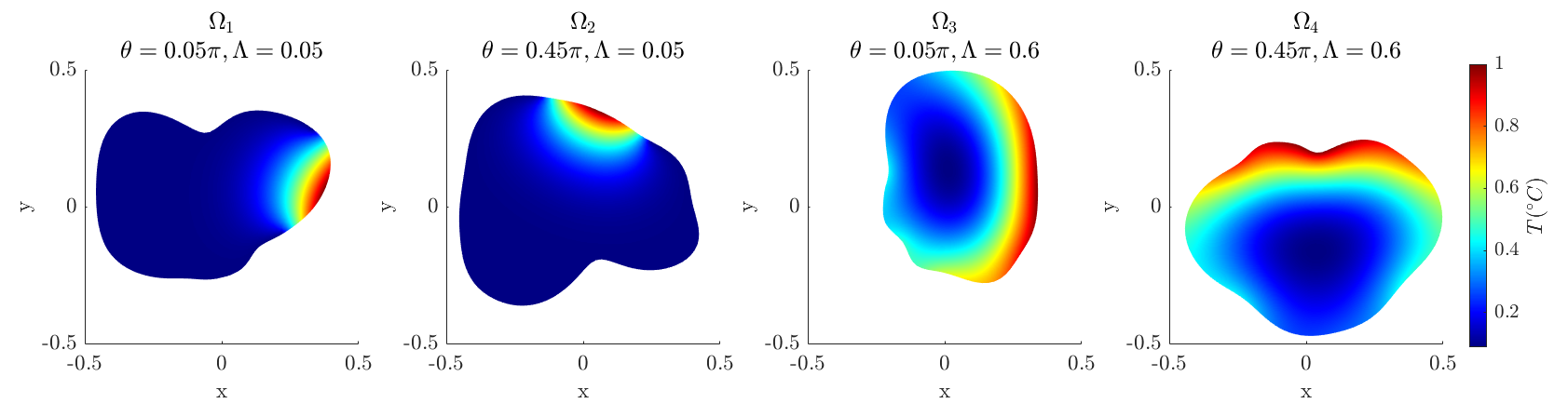} 
	\caption{Sample training solutions at four parameter vectors sampled in $\mathcal{D}$ -- from left to right: $\Omega^{1}$; $\Omega^{2}$; $\Omega^{3}$; and $\Omega^{4}$.} 
	\label{fig:Solutions} 
\end{figure}

For each geometry, $P=30$ solutions of the heat transfer problem are computed across the parameter domain $\mathcal{D} = (\theta, \Lambda)$. The design of experiments (DoE) utilized 
for each geometry is uniform and follows a latin hypercube sampling (LHS) method. The parameters explore ranges of $[0.05 \pi, 0.45 \pi]$ for $\theta$ and $[0.05, 0.6]$ for $\Lambda$.
These solutions are obtained in each case using a finite element analysis (FEA) with a fine mesh, as illustrated in Figure \ref{fig:Geometries_mesh}, which shows the fine meshes used in the computations. 
Figure \ref{fig:Solutions} presents examples of the computed solutions, showcasing one for each randomly generated geometry at a different sampled parameter vector, resulting in 
varying levels of solution localization.

The four randomly generated geometrical domains $\Omega^{k}$ $(k \in \llbracket 4 \rrbracket)$ illustrated in Figure \ref{fig:Geometries}, along with the previously described DoE 
using $\mathcal{D} = (\theta, \Lambda)$ and resulting in 120 solution snapshots $(K \times P = 4 \times 30 = 120)$, will be used in the remainder of this paper to illustrate
the various components of the proposed OT-based interpolation methodologies for supporting the GD vision outlined in Section \ref{sec:INTRO}.

\subsection{Complexity of the Overall Problem}

Although the chosen parametric transient heat transfer problem involves relatively simple and well-understood physics, the associated 
geometry and data interpolation tasks are deliberately challenging. In particular, the proposed OT-based interpolation framework must 
interpolate solution snapshots across randomly generated, arbitrarily shaped domains $\Omega^k$, each discretized using one of four 
topologically distinct meshes of varying sizes -- a task that is notoriously difficult for alternative scientific machine learning 
approaches, such as projection-based model order reduction. Additionally, the parameterized boundary conditions produce 
temperature fields with strongly localized features -- steep thermal gradients and boundary-layer structures that translate and 
deform across the design space -- posing a stringent test for any interpolation method. Finally, constructing the surrogate requires 
solving a $K \times P$-dimensional matching problem (for example, $4$ geometries $\times$ $30$ snapshots $=120$ total) to establish 
particle correspondences across all domains and parameter instances, illustrating the intrinsic complexity of preprocessing real 
multi-parametric, multi-geometry systems.

\subsection{Particle Decomposition}

We begin by determining suitable values for the hyperparameters $N_{s}$ and $\sigma_{s}$, which govern the decomposition of a solution snapshot into a sum of Gaussian functions. 
These parameters significantly influence the particles' ability to capture complex fields that may exhibit localized features. To achieve this, we conduct a hyperparameter study 
across three distinct ranges of the DoE along the $\Lambda$ dimension of $\mathcal{D}$: specifically, $\Lambda \in [0.05, 0.2]$, $\Lambda \in [0.2, 0.4]$, and $\Lambda \in 
[0.4, 0.6]$, at each of the $K=4$ randomly generated geometries.

\begin{figure}[H]
    \centering
    \begin{subfigure}{0.27\textwidth}
        \includegraphics[width=\textwidth]{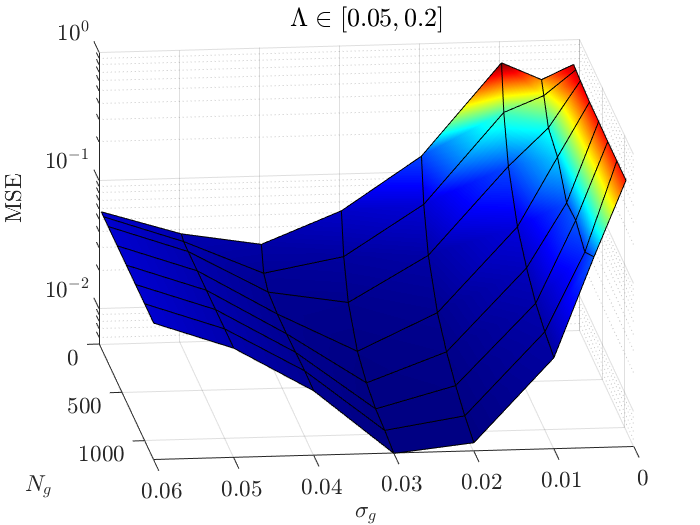} 
	    \caption{First $\Lambda$ interval.} 
	    \label{fig:SPH_0.05_0.2} 
    \end{subfigure}
    \hfill
    \begin{subfigure}{0.27\textwidth}
        \includegraphics[width=\textwidth]{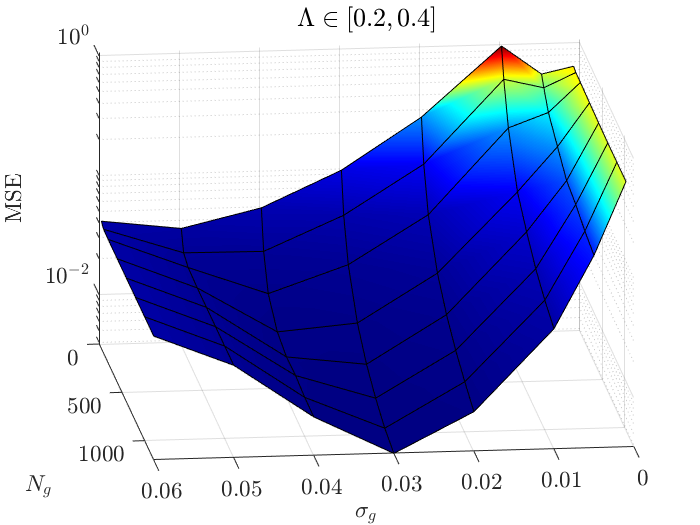} 
	    \caption{Second $\Lambda$ interval.} 
	    \label{fig:SPH_0.2_0.4}
    \end{subfigure}
    \hfill
    \begin{subfigure}{0.27\textwidth}
        \includegraphics[width=\textwidth]{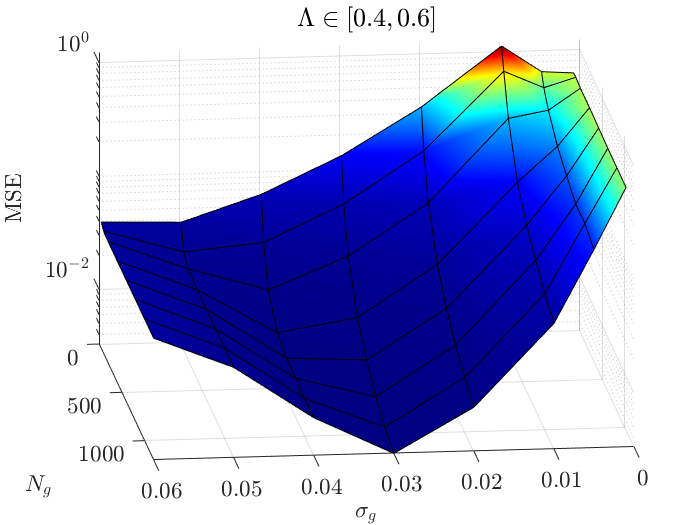}
	    \caption{Third $\Lambda$ interval.} 
	    \label{fig:SPH_0.4_0.6}
    \end{subfigure} 
	\caption{Error analysis of particle decomposition across all four randomly generated geometries and three intervals of $\Lambda$.} 
	\label{fig:SPH_analysis} 
\end{figure}

For each range, we focus on the mean parameter vector of the associated reduced DoE and the solution snapshot computed at this vector, which defines a specific boundary condition. 
We evaluate how effectively this solution snapshot can be represented as a sum of $N_{s}$ Gaussian functions with a fixed standard deviation $\sigma_{s}$ by testing various values 
for both $N_{s}$ and $\sigma_{s}$. Across all $K=4$ considered geometries, we compute the mean squared error (MSE) of the representation for each case, as illustrated in 
Figure \ref{fig:SPH_analysis}.

Based on this analysis, the empirically determined near-optimal combination of hyperparameters is $N_{s} = 600$ and $\sigma_{s} = 0.03$, which will be used in all subsequent 
evaluations. 

\begin{figure}[H] 
	\centering 
	\includegraphics[width=1\textwidth]{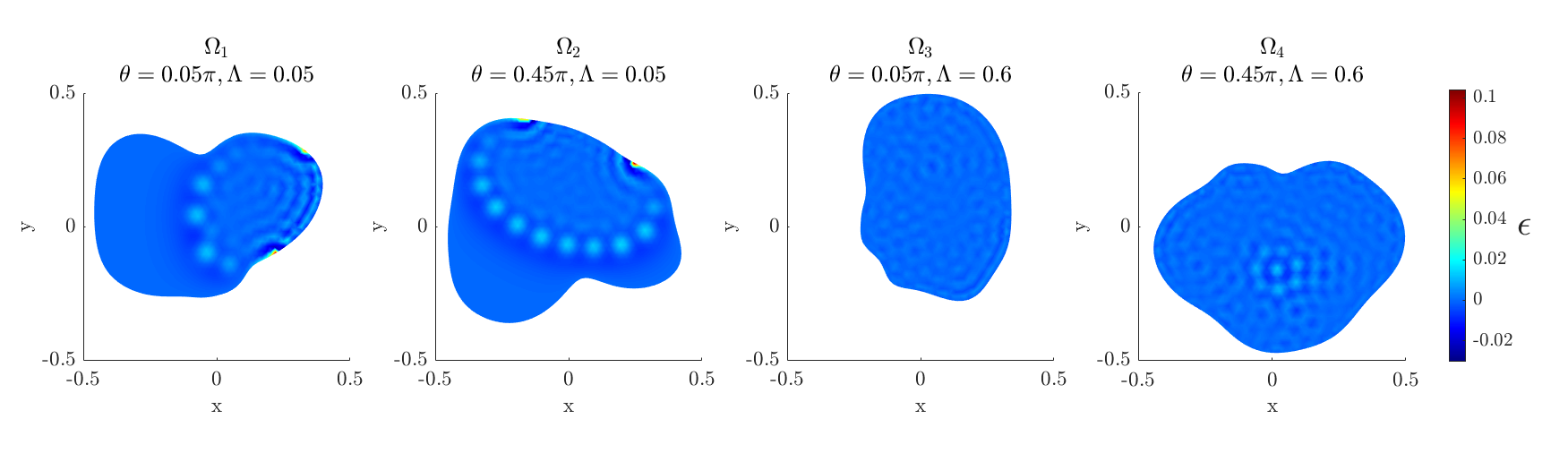} 
	\caption{Relative error $\epsilon$ between the original and reconstructed solutions for each randomly generated geometry ($\Omega^{1}$ to $\Omega^{4}$, from left to right), 
	with the training parameter vector indicated for each geometry.}
	\label{fig:SPH1}
\end{figure}

\begin{figure}[H] 
	\centering 
	\includegraphics[width=1\textwidth]{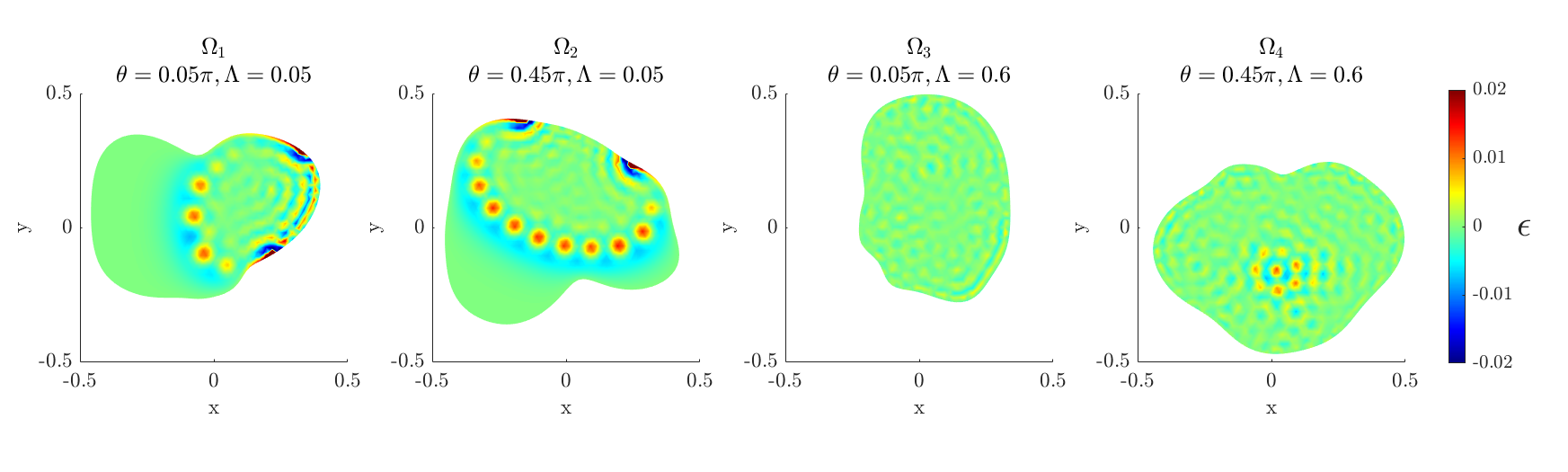} 
	\caption{Relative error $\epsilon$ between the original and reconstructed solutions for each randomly generated geometry ($\Omega^{1}$ to $\Omega^{4}$, from left to right), 
	shown with a fixed colorbar ranging from $-0.02$ to $0.02$, and highlighting the training parameter vector for each geometry.}
	\label{fig:SPH2} 
\end{figure}

Then, we test the decomposition into Gaussian particles by comparing the original and reconstructed solutions of problem \eqref{eq:heat} for each of the $K=4$ randomly generated 
geometries at various parameter vectors in $\mathcal{D}$. For this purpose, we define the following relative error $\epsilon$ between the original $\psi$ and reconstructed 
$\hat{\psi}$ solutions
\begin{equation}
	\epsilon = \displaystyle{\frac{\hat{\psi} - \psi}{\sqrt{\int_{\Omega} \frac{\psi^{2}}{\mathcal{A}_{\Omega}} \dd \Omega}}}
	\label{eq:RE}
\end{equation}
where $\mathcal{A}_{\Omega}$ is the surface area of the domain $\Omega$. The resulting errors are graphically depicted in Figures \ref{fig:SPH1} and \ref{fig:SPH2}, highlighting the 
training parameter vector for each geometry. These figures collectively show that, with the exception of two very small regions in $\Omega^{1}$ and $\Omega^{2}$ where the relative 
error peaks at 10\%, the relative error generally remains below 2\%. This very small error confirms the near-optimal choice of the aforementioned hyperparameters.

\subsection{Illustration of the Surrogate Solution Model for a Specific Geometry}

We now focus on the geometrical domain $\Omega^1$ (shown in Figure \ref{fig:Geometries}-left) and its corresponding mesh (illustrated in Figure \ref{fig:Geometries_mesh}-left). For 
this domain, we construct the SSM $\mathcal{S}^1$ to illustrate the OT-based methodology for interpolating parametric fields associated with a specific geometry, as described in this 
paper.

\begin{figure}[H] 
	\centering 
	\includegraphics[width=1\textwidth]{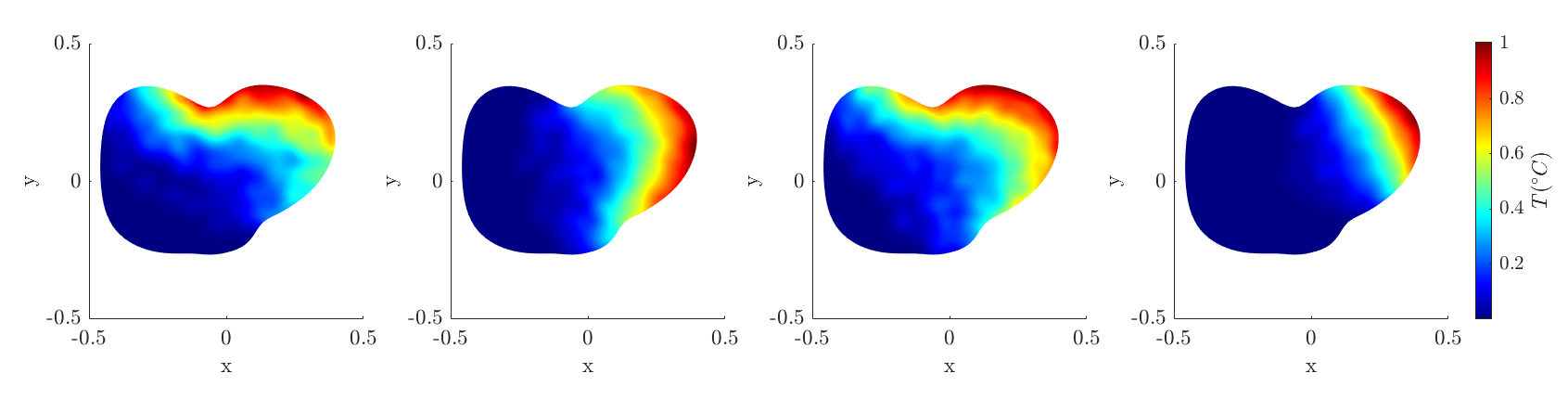} 
	\caption{Inferred heat transfer solutions in $\Omega^{1}$ at four distinct parameter vectors in $\mathcal{D}$.} 
	\label{fig:Solution_FixedGeometry_Infer} 
\end{figure}

\begin{figure}[H] 
	\centering 
	\includegraphics[width=1\textwidth]{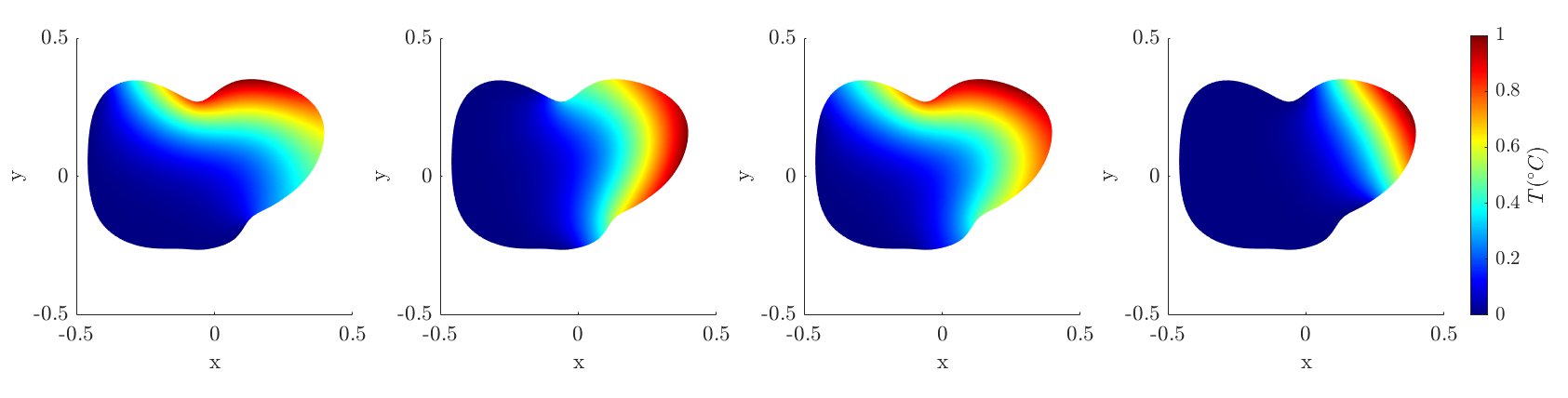} 
	\caption{FEA heat transfer reference solutions in $\Omega^{1}$ at four distinct parameter vectors in $\mathcal{D}$.} 
	\label{fig:Solution_FixedGeometry_Ref} 
\end{figure}

\begin{figure}[H] 
	\centering 
	\includegraphics[width=1\textwidth]{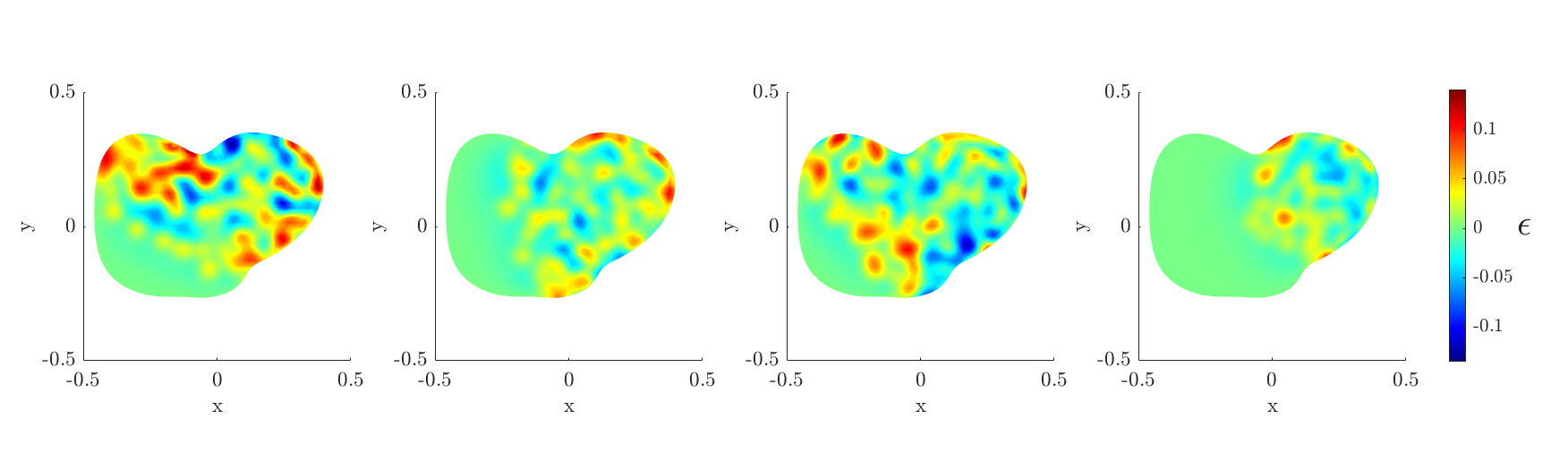} 
	\caption{Relative error $\epsilon$ between each inferred solution $T$ and the corresponding FEA reference solution $T_\text{ref}$.}
	\label{fig:Solution_FixedGeometry_Error} 
\end{figure}

Specifically, four parameter vectors $(\theta_i,\Lambda_i) \in \mathcal D$ for $i \in \llbracket 4 \rrbracket$ are considered, each defining a different boundary condition for the 
heat transfer problem \eqref{eq:heat}. At each parameter vector, the solution within $\Omega^{1}$ is inferred, as presented in Figure \ref{fig:Solution_FixedGeometry_Infer}. 
Additionally, the corresponding heat transfer problems are solved using FEA with the same material properties used for generating the training snapshots; these FEA reference solutions
are shown in Figure \ref{fig:Solution_FixedGeometry_Ref}. Finally, the relative error $\epsilon$ \eqref{eq:RE} between the inferred solution $T$ and the corresponding FEA reference 
solution $T_\text{ref}$ is computed and depicted in Figure \ref{fig:Solution_FixedGeometry_Error}.

\begin{figure}[H] 
	\centering \includegraphics[width=0.8\textwidth]{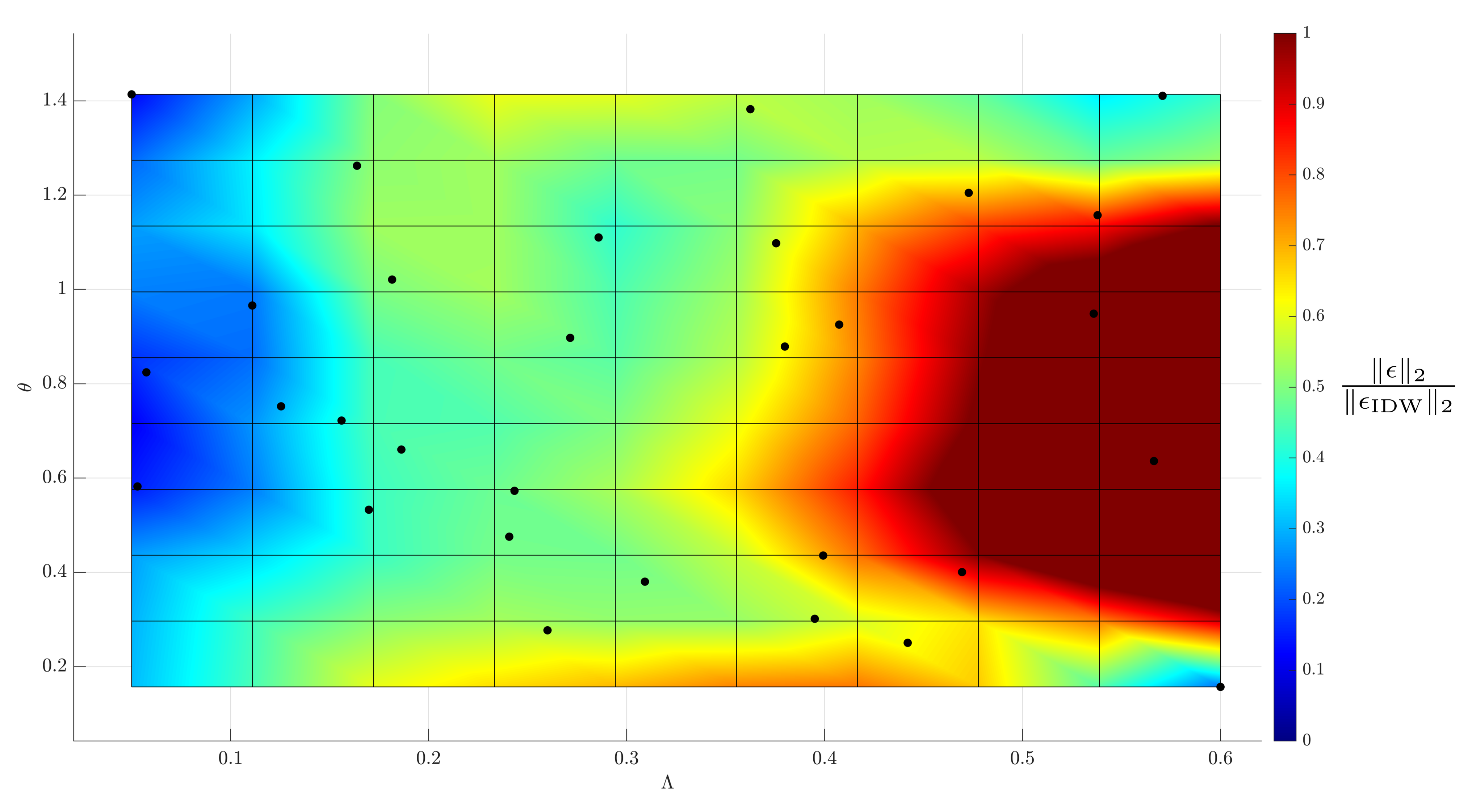} 
	\caption{Quotient of the Euclidean norms of the relative errors $\epsilon$ and $\epsilon_\text{IDW}$, with black dots indicating the training points.} 
	\label{fig:ErrorSurrogate_ErrorRBF} 
\end{figure}

\begin{figure}[H] 
	\centering 
	\includegraphics[width=0.8\textwidth]{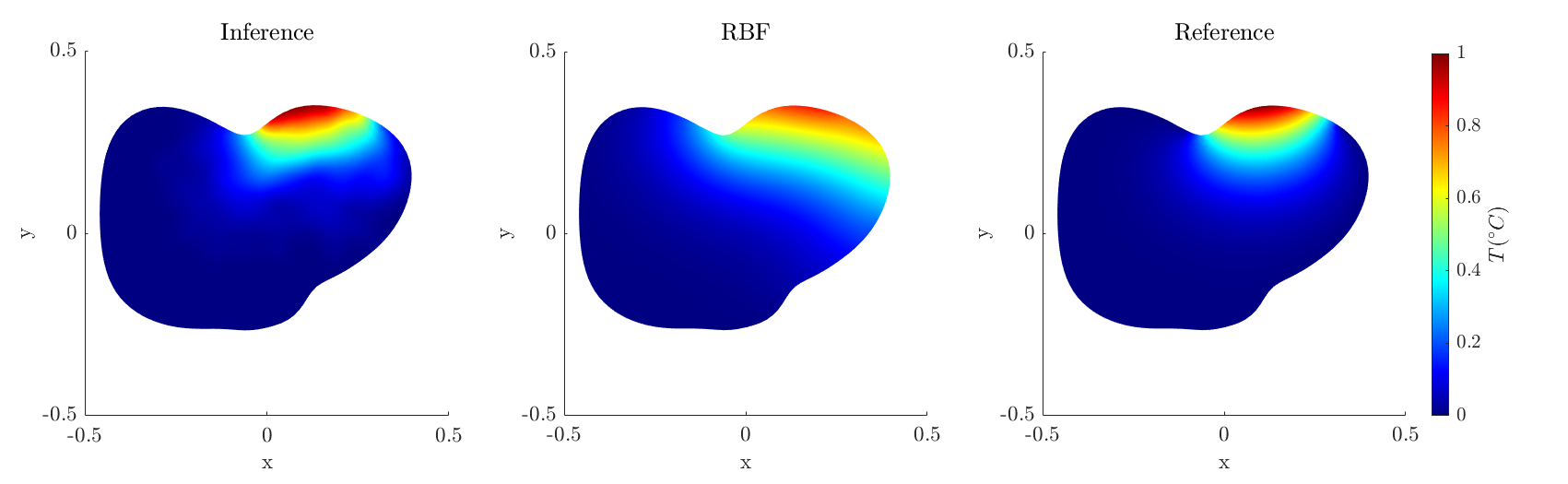} 
	\caption{Comparison of a localized field: OT-based inferred solution (left), IDW interpolated solution (middle), and FEA reference solution (right).}
	\label{fig:Surrogate_RBF} 
\end{figure}

To assess performance, we now compare the relative error $\epsilon$ \eqref{eq:RE} with the relative error $\epsilon_\text{IDW}$ obtained from an inverse distance weighting (IDW) 
interpolation method. In this IDW interpolation, the weights of neighbor solutions are computed using a radial basis function in the parameter space. This comparison, depicted in 
Figure \ref{fig:ErrorSurrogate_ErrorRBF}, clearly shows that the surrogate method excels for the most localized fields (small values of $\Lambda$, see Figure \ref{fig:Solutions}). In 
such cases, the IDW interpolation yields unsatisfactory results, as further illustrated in Figure \ref{fig:Surrogate_RBF}.

\subsection{Inferring Snapshots With Surrogate Geometry and Solution Models}

\begin{figure}[H] 
	\centering 
	\begin{subfigure}{0.455\textwidth} 
		\includegraphics[width=\textwidth]{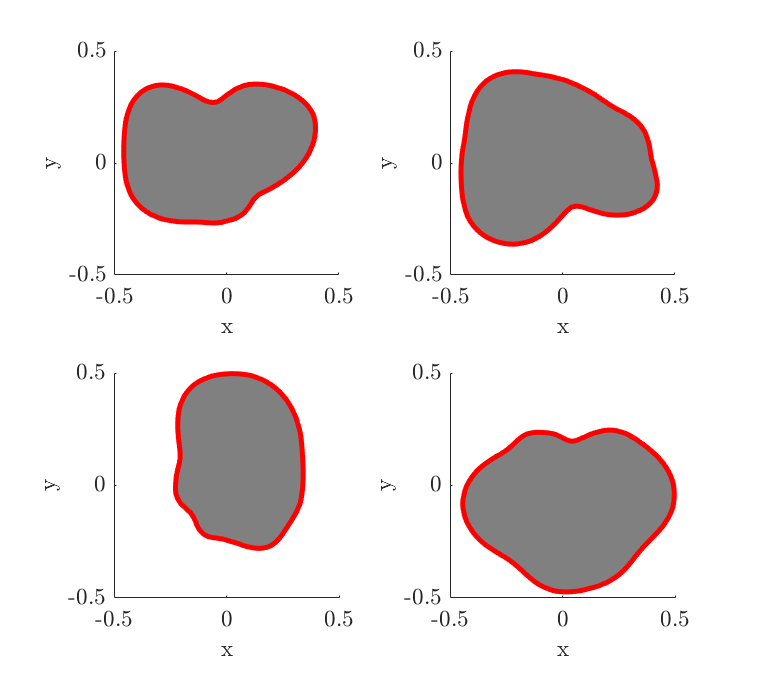} 
		\caption{Original geometry outlined in red, with the reconstructed geometry $\hat{\varphi}^{k}$ highlighted in grey.}
		\label{fig:Heat_geometries} 
	\end{subfigure} 
	\hfill 
	\begin{subfigure}{0.465\textwidth} 
		\includegraphics[width=\textwidth]{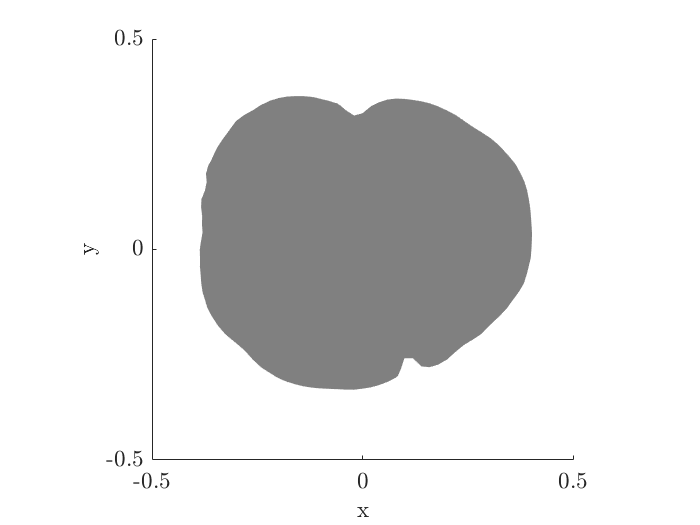} 
		\caption{Interpolated geometry $\hat{\varphi}^{\star}$ with equal weights $\omega^{1}$ = $\omega^{2}$ = $\omega^{3}$ = $\omega^{4}$ = 0.25.} 
		\label{fig:Heat_geometries_New} 
	\end{subfigure}		
	\label{fig:Heat_geometries_SPH} 
	\caption{Reparameterization of each of $\Omega^1$ (top-left), $\Omega^2$ (top-right), $\Omega^3$ (bottom-left), and $\Omega^4$ (bottom-right) from a signed distance function 
	to a sum of Gaussian functions; interpolation of the reparameterizations of $\Omega^{1}$ to $\Omega^{4}$ using specified weights to generate $\Omega^{\star}$ (bottom).}
\end{figure}

Following the empirical determination of the near-optimal hyperparameters $N_g = 600$ and $\sigma_g = 0.02$, which will be consistently applied in all subsequent 
evaluations, we proceed to reconstruct each of the randomly generated geometrical domains $\Omega^k$ for $k \in \llbracket 4 \rrbracket$ with a set of particles, as 
described in Section \ref{sec:GENINTERP}. In Figure \ref{fig:Heat_geometries}, we compare the reconstructed geometries $\hat{\varphi}_{k}$ for $k \in \llbracket 4 \rrbracket$ 
with their original counterparts, demonstrating that the chosen hyperparameters $N_g$ and $\sigma_g$ are suitably calibrated. Furthermore, Figure \ref{fig:Heat_geometries_New} 
illustrates the geometry $\Omega^{\star}$, created using these hyperparameters and equal weights $\omega^{1}$ = $\omega^{2}$ = $\omega^{3}$ = $\omega^{4}$ = 0.25.
	
\begin{figure}[H] 
	\centering 
	\includegraphics[width=1.0\textwidth]{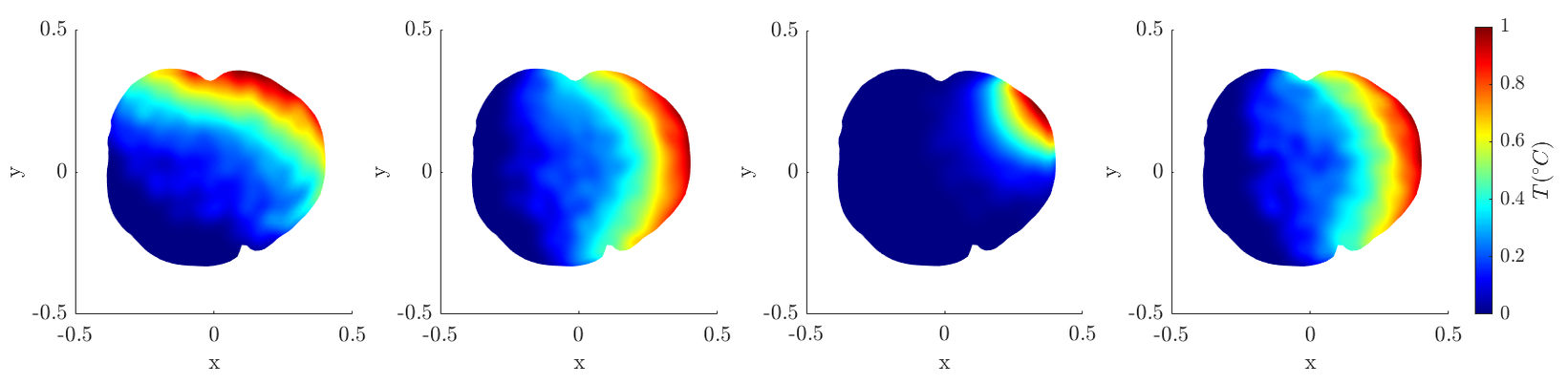} 
	\caption{Four inferred heat transfer solution snapshots in the geometrical domain $\Omega^{\star}$.} 
	\label{fig:Heat_geometries_Infer} 
\end{figure}

\begin{figure}[H] 
	\centering 
	\includegraphics[width=1.0\textwidth]{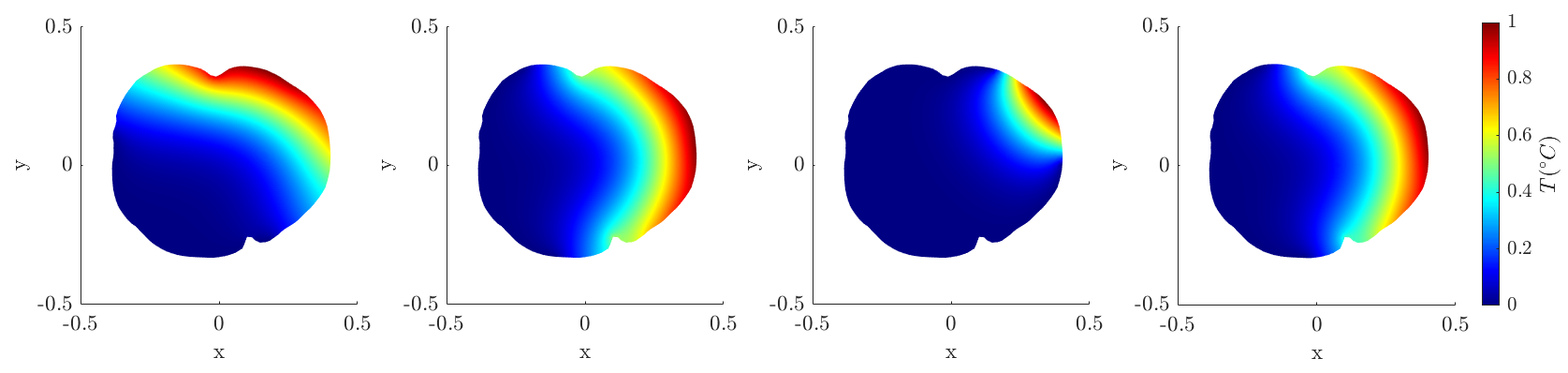} 
	\caption{Four FEA heat transfer solution snapshots in the geometrical domain $\Omega^{\star}$.} 
	\label{fig:Heat_geometries_Ref}
\end{figure}

\begin{figure}[H] 
	\centering 
	\includegraphics[width=1.0\textwidth]{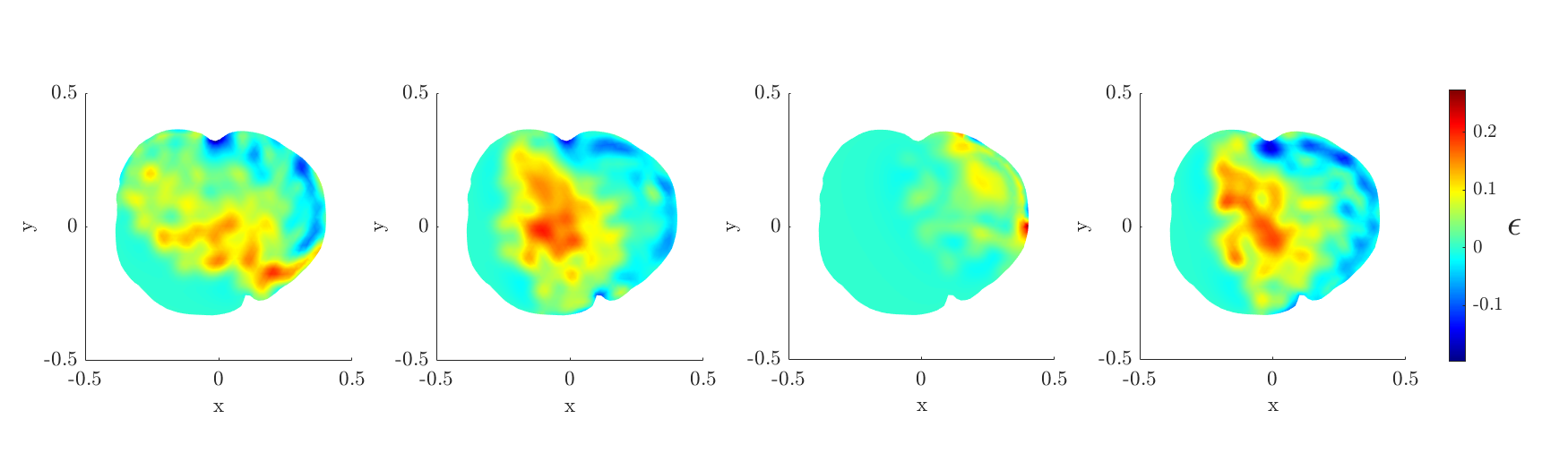} 
	\caption{Relative error $\epsilon$ between the four inferred solution snapshots $T$ and the corresponding four reference FEA solution snapshots $T_{ref}$.} 
	\label{fig:Heat_geometries_Error} 
\end{figure}

We construct a SSM $\mathcal{S}^k$ for each geometrical domain $\Omega^k$, with $k \in \llbracket 4 \rrbracket$. Each SSM is used to 
demonstrate the OT-based solution surrogate model -- an overview of which is shown in Figure \ref{fig:OToverview_geo} -- on a 
reconstructed geometry. To achieve this, we first interpolate each set of $P=30$ solutions from the training 
snapshots associated with each domain onto the target geometrical domain $\Omega^\star$. This interpolation is performed with the same equal weights ($\omega^{1}$ = $\omega^{2}$ = 
$\omega^{3}$ = $\omega^{4}$ = 0.25) that were also applied in the context of the reconstructed geometrical domain $\Omega^\star$. A subset of four of these resulting interpolated 
snapshots is visualized in Figure \ref{fig:Heat_geometries_Infer}. Additionally, the geometrical domain $\Omega^{\star}$ is discretized with a mesh, and the 120 reference solutions of
the DoE are computed using FEA. Four of these FEA-computed reference solutions, corresponding to the interpolated ones, are illustrated in Figure \ref{fig:Heat_geometries_Ref}. 
The relative error $\epsilon$ \eqref{eq:RE} between these two sets of solutions is also computed and displayed in Figure \ref{fig:Heat_geometries_Error}. This error peaks at 20\%, 
but only in very small regions of the computational domains where the temperature field is nearly uniform and exhibits its lowest values.

\begin{figure}[H] 
	\centering 
	\includegraphics[width=1.0\textwidth]{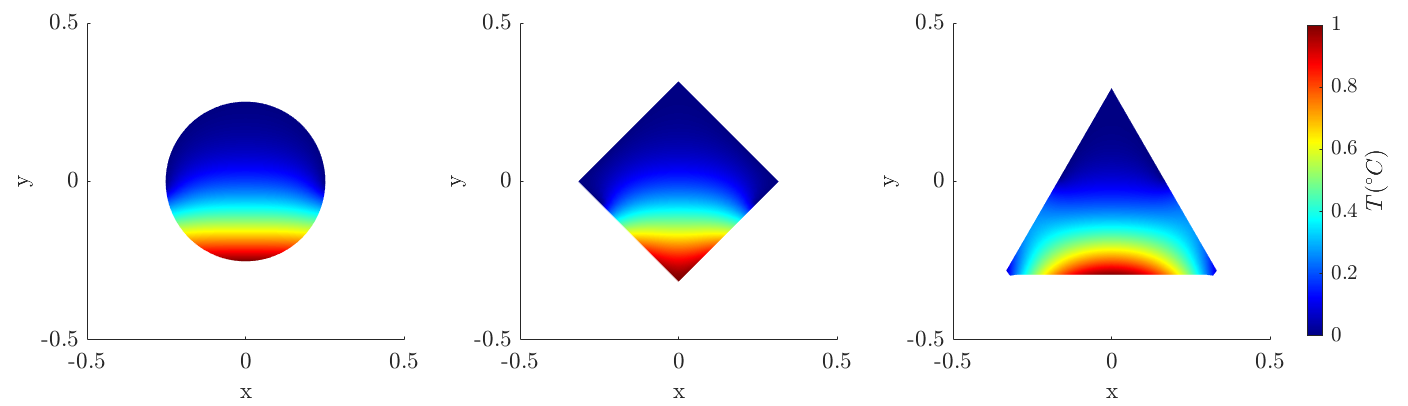} 
	\caption{FEA heat transfer solution snapshots in three different geometries.} 
	\label{fig:TCS_Sol} 
\end{figure}

\begin{figure}[H] 
	\centering 
	\includegraphics[width=1.0\textwidth]{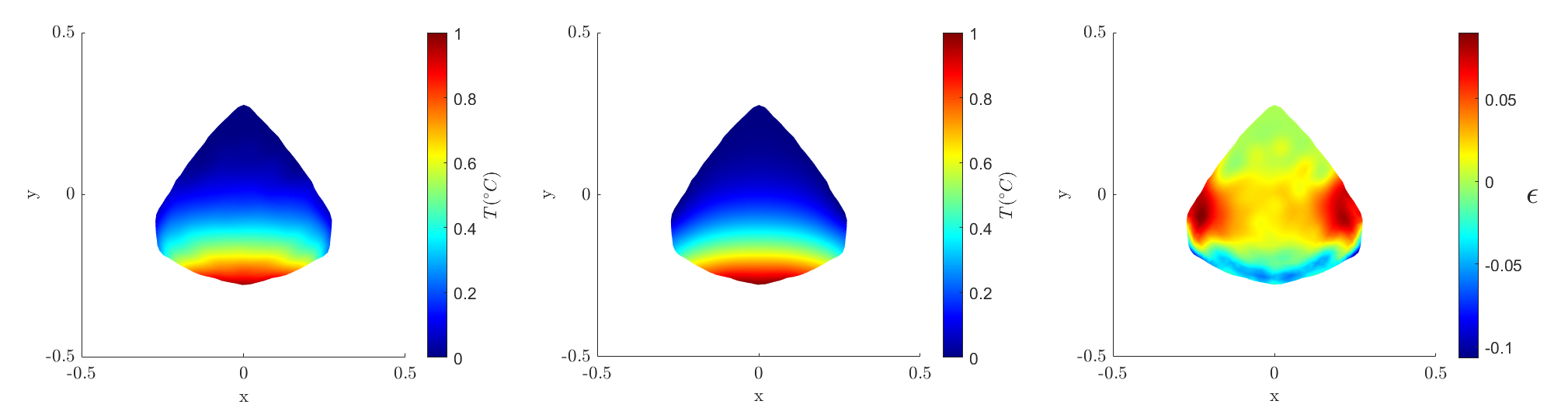} 
	\caption{Inferred (left) and FEA (middle) heat transfer solution snapshots in the interpolated geometrical domain $\Omega^{\star}$, along with the corresponding relative
	error $\epsilon$ (right).}
	\label{fig:TCS_Error} 
\end{figure}

We further illustrate the robustness of the OT-based geometry interpolation methodology on a less regular problem. We begin by considering three distinct geometries: a circle, a 
diamond, and a triangle, along with heat transfer solutions of problem \eqref{eq:heat} computed using FEA for these geometries (see Figure \ref{fig:TCS_Sol}). We then reconstruct 
these geometries and interpolate them, assigning weights of $\omega^{1}$ = 0.3, $\omega^{2}$ = 0.3, and $\omega^{3}$ = 0.4, to derive the geometrical domain $\Omega^\star$ in 
reconstruction form, shown in Figure \ref{fig:TCS_Error}. After meshing $\Omega^\star$, we compute a FEA heat transfer reference solution within it. We proceed to interpolate the 
FEA reference solutions from the circle, diamond, and triangle within $\Omega^\star$ and calculate the relative error $\epsilon$ \eqref{eq:RE} between this interpolated solution and 
its FEA counterpart. The results are presented in Figure \ref{fig:TCS_Error}.
	
\subsection{Performing Predictions With Surrogate Geometry and Solution Models}

In this section, we illustrate the application of the combined OT-based interpolation methodologies presented in this paper to perform predictions within a ``what-if'' context 
pertinent to GD.

Our initial step involves empirically determining the near-optimal value of the sole hyperparameter of the sPGD regressor: the degree of the polynomial approximation it relies upon. 
We remind the reader that this sPGD regressor is utilized in the Regressor Training substep (detailed in Section \ref{sec:OFFLINE} of the offline stage within the OT-based 
interpolation methodology described in Section \ref{sec:BACKPLUS}) to infer the $N_s$ particle positions for any parameter vector $\bm{\theta}^\star \in \mathcal{D}$ not included in 
the training set.

\begin{figure}[H] 
	\centering 
	\includegraphics[width=0.7\textwidth]{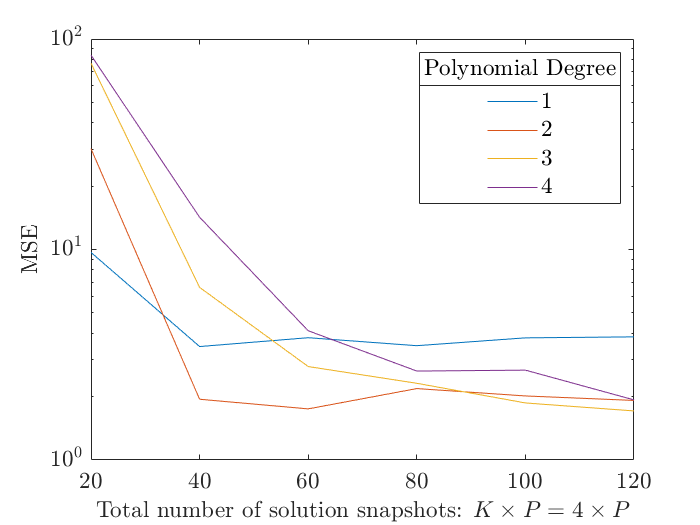} 
	\caption{Regression mean squared error as a function of the total number of snapshots and polynomial degree used in sPGD.}
	\label{fig:Method_Error}
\end{figure}

To optimize this hyperparameter, we train the regressor using a varying number of solution snapshots from the DoE (ranging from 20 to 120) and polynomial degrees for sPGD 
(from 1 to 4). The resulting regression MSE is reported in Figure \ref{fig:Method_Error}, based on which we select a polynomial approximation of degree 3 for the sPGD regressor.

\begin{figure}[H] 
	\centering 
	\includegraphics[width=1.0\textwidth]{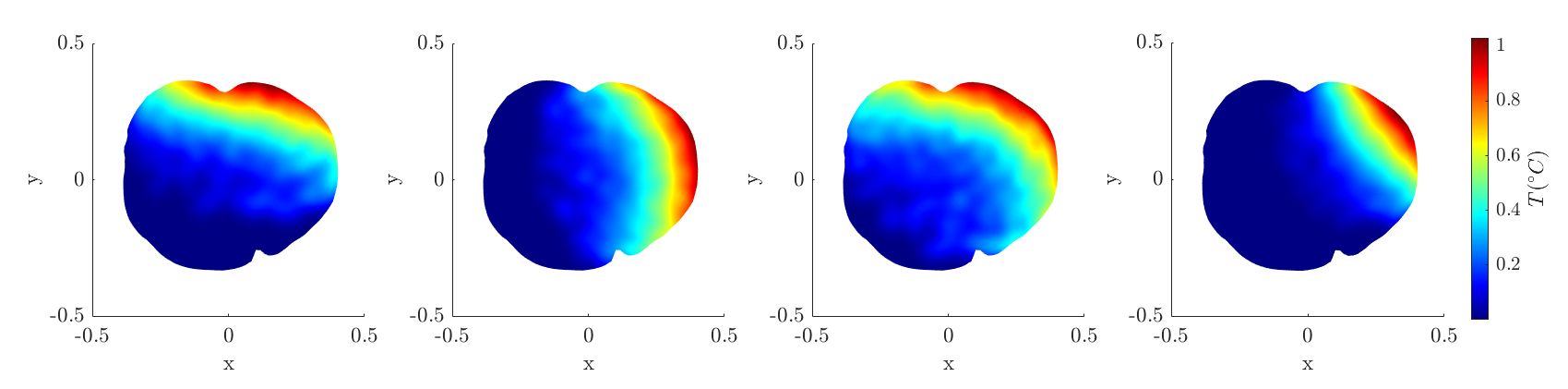} 
	\caption{Predicted heat transfer solutions in $\Omega^\star$ at four queried parameter vectors not included in the training set.}
	\label{fig:Method_Infer} 
\end{figure}

\begin{figure}[H] 
	\centering 
	\includegraphics[width=1.0\textwidth]{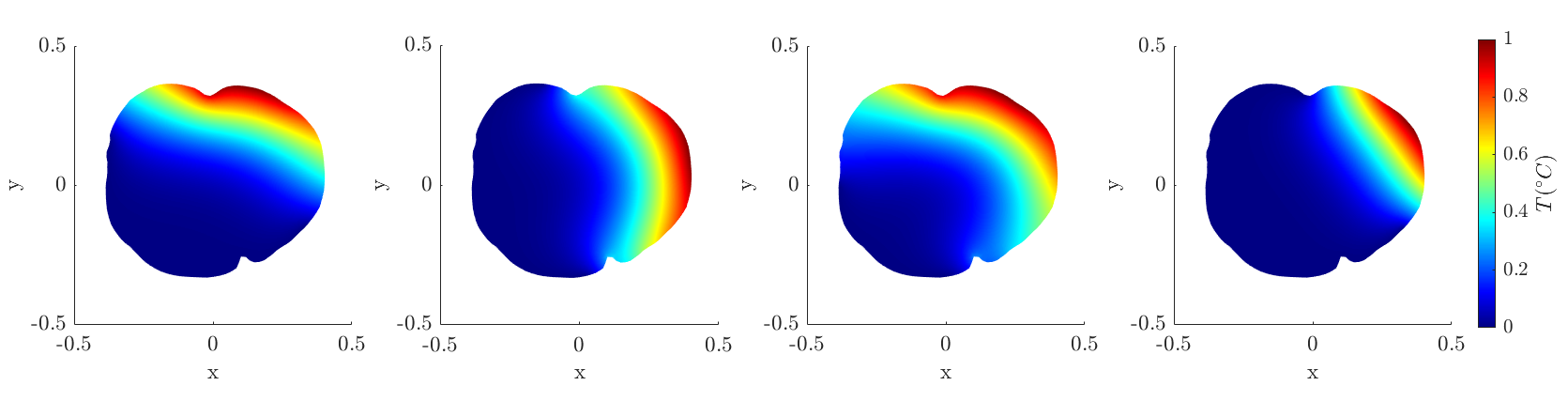} 
	\caption{FEA heat transfer reference solutions in $\Omega^{\star}$ at four queried parameter vectors not included in the training set.}
	\label{fig:Method_Ref} 
\end{figure}

\begin{figure}[H]  
	\centering 
	\includegraphics[width=1.0\textwidth]{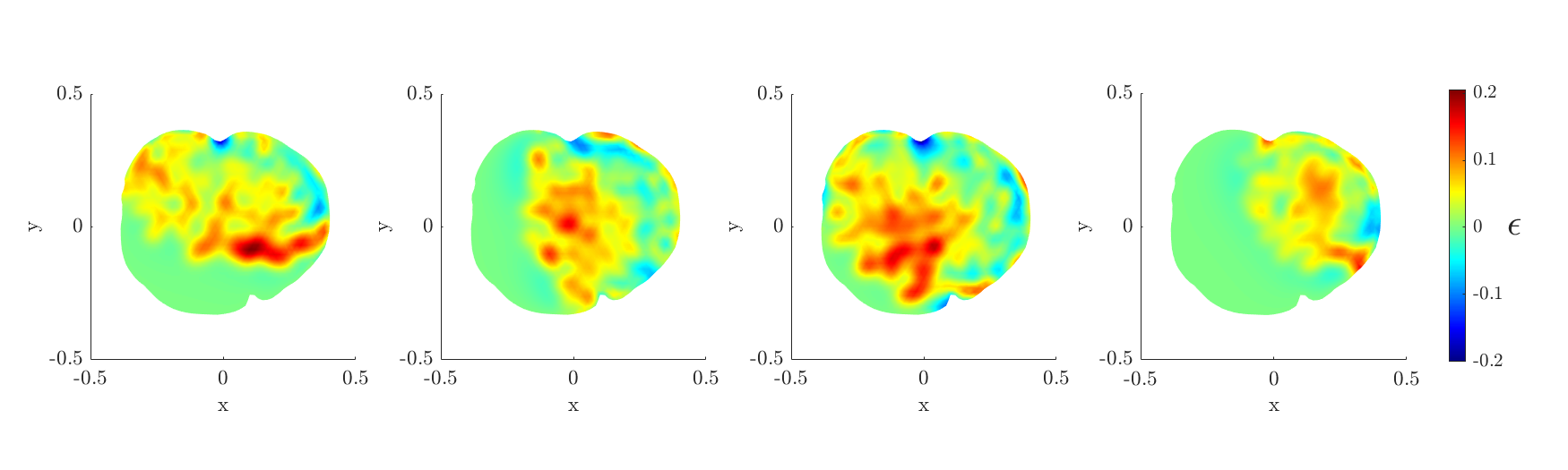} 
	\caption{Relative error $\epsilon$ between the predicted and reference FEA solutions evaluated at four queried parameter vectors not included in the training set.}
	\label{fig:Surrogate_analysis} 
\end{figure}

Next, we sample four test parameter vectors within the parameter domain $\mathcal{D} = (\theta, \Lambda)$, ensuring these vectors were not part of the DoE. Using the four SSMs 
$\mathcal{S}^k$ introduced in the previous section ($k \in \llbracket 4 \rrbracket$), we infer the corresponding solutions within the randomly generated geometrical domains 
$\Omega^k$ ($k \in \llbracket 4 \rrbracket$). Subsequently, we interpolate a new geometrical domain $\Omega^\star$ by assigning equal weights ($\omega^1$ = $\omega^2$ = $\omega^3$ 
= $\omega^4$ = 0.25) to the four original domains. We also interpolate the inferred solutions from these four geometrical domains onto $\Omega^\star$, using the same equal weights. 
The resulting four interpolated solutions, their corresponding FEA reference solutions, and the associated relative errors $\epsilon$ \eqref{eq:RE} are presented in Figure 
\ref{fig:Method_Infer}, Figure \ref{fig:Method_Ref}, and Figure \ref{fig:Surrogate_analysis}, respectively. 

As evidenced by Figures \ref{fig:Method_Infer} to \ref{fig:Surrogate_analysis}, the heat transfer solutions predicted for several 
out-of-sample parameter vectors 
exhibit residual errors that are mostly small relative to the overall temperature field. Furthermore, the error observed in certain regions of the interpolated geometrical domain
$\Omega$ is comparable to the scale of individual particles in the field decomposition. These findings highlight opportunities for enhancement through a more refined choice of 
hyperparameters and underscore the importance of developing an automated hyperparameter setting method.

\subsection{Execution Time Performance}

Finally, we evaluate the performance of our combined OT-based interpolation methodologies on the parametric heat transfer problem \eqref{eq:heat} using the same $K=4$ randomly 
generated geometries shown in Figure \ref{fig:Geometries}. For this evaluation, we utilize the previously established near-optimal hyperparameters: $N_s = 600$, $N_g = 600$, 
$\sigma_s = 0.03$, and $\sigma_g = 0.02$. Although this example is computationally inexpensive and not the most realistic, it clearly demonstrates certain scalability aspects of the 
offline performance of the proposed OT-based interpolation methodologies and their online performance. For larger-scale computational applications, the overall efficiency gains would 
be far more significant, particularly for the online performance.

In the context of the parametric two-dimensional transient heat transfer problem examined in this work, the computational expense of the offline stage of our combined OT-based 
interpolation methodologies is predominantly attributed to the execution of the DoE and the solution of the $K \times P$-dimensional matching problem. Employing a mesh 
comprising 2\,220 elements and 1\,200 nodes (depending on the geometry), each implicit FEA solution of the heat transfer problem \eqref{eq:heat} incurs a computational time of 
0.22 seconds on the single CPU core detailed at the beginning of Section \ref{sec:APP}. Given the linear scaling of the DoE cost with the number of snapshots, the generation of 
120 snapshots necessitates 26.4 seconds of wall clock time.

\begin{figure}[H] 
	\centering 
	\includegraphics[width=0.7\textwidth]{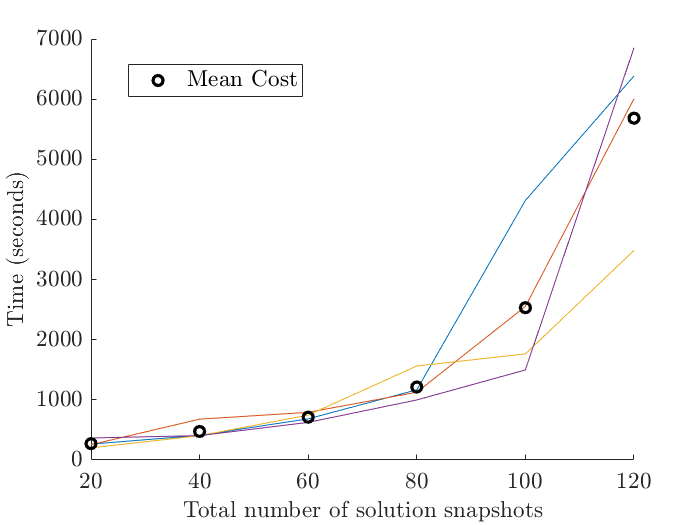} 
	\caption{Influence of the number of solution snapshots on the computational cost (wall clock time) of solving the $K \times P$-dimensional matching problem using a genetic 
	algorithm.}
	\label{fig:GA_analysis}
\end{figure}

To investigate the influence of the total number of collected solution snapshots on the computational cost of the $K \times P$-dimensional matching problem, for each of the considered
four geometries ($K=4$), we systematically increase the number of snapshots from 5 to 30 in steps of 5, consequently varying the total number of snapshots $K \times P$ between 20 and 
120. To assess how this variation affects the $K \times P$-dimensional matching problem, we record the computational cost (in seconds) of the genetic algorithm used for its solution,
maintaining a consistent convergence criterion throughout this investigation. This process is repeated several times on the same single-processor computing system, and the average 
computational cost as a function of the total number of solution snapshots is illustrated in Figure \ref{fig:GA_analysis}. 
As shown in this figure, the computational cost of the matching problem increases roughly linearly for 20 to 80 total snapshots, 
followed by an almost exponential rise beyond 80, consistent with the exponential growth of the MAP complexity as the number of 
dimensions $P$ increases.

To investigate how the total number of collected solution snapshots affects the computational cost of the $K \times P$-dimensional matching problem, we systematically increase the 
number of snapshots per geometry (for the four considered) from 5 to 30 in steps of 5, resulting in a total snapshot count ranging from 20 to 120. For each total, we record the 
computational cost (in seconds) of the genetic algorithm used to solve the matching problem, maintaining a consistent convergence criterion. This process is repeated multiple times 
on the same single-processor system, and the average computational cost as a function of the total number of solution snapshots is shown in Figure \ref{fig:GA_analysis}. The figure 
reveals a roughly linear increase in computational cost between 20 and 80 total snapshots, followed by a near-exponential rise beyond 80.

In conclusion, the cumulative offline computational cost (wall clock time) for our integrated OT-based interpolation methodologies, considering a DoE comprising 120 snapshots, 
amounts to 6\,530.3 seconds (see Tables \ref{tab:CP1} and \ref{tab:CP2} for further details). However, the online inference of a solution for the studied parametric two-dimensional 
transient heat transfer problem at an out-of-sample parameter vector within a geometrical domain $\Omega^\star$, interpolated using an SGM and four SSMs, is achieved in a mere 0.02 seconds. 
This is an order of magnitude faster than a numerical prediction based on FEA, and this speedup factor will increase dramatically for larger-scale applications and longer simulation 
time intervals.

\begin{table}[H]
	\centering
	\begin{tabular}{|cccc|}
		\cline{1-4}
		\multicolumn{4}{|c|}{\textbf{SSM Offline Stage}}   \\ \cline{1-4}
		\multicolumn{1}{|c|}{Particle Decomposition}      & \multicolumn{1}{c|}{4.5}  & \multicolumn{1}{c|}{$\times 120$} & 541.7    \\ \cline{1-4}
		\multicolumn{1}{|c|}{$P$-Dimensional Matching ($P$ = 120)} & \multicolumn{2}{c|}{}                                               & 5\,683.2        \\ \cline{1-4}
		\multicolumn{1}{|c|}{SSM Training}    & \multicolumn{1}{c|}{2.0}  & \multicolumn{1}{c|}{$\times 4$}   & 8.2      \\ \cline{1-4}
		\multicolumn{4}{|c|}{\textbf{SGM Offline Stage}}   \\ \cline{1-4}
		\multicolumn{1}{|c|}{Particle Decomposition}      & \multicolumn{1}{c|}{56.1} & \multicolumn{1}{c|}{$\times 4$}   & 224.5   \\ \cline{1-4}
		\multicolumn{1}{|c|}{$K$-Dimensional Matching ($K$ = 4)}   & \multicolumn{2}{c|}{}                                               & 80.9     \\ \cline{1-4}
		\multicolumn{3}{|c|}{\textbf{Offline Total}}   & \textbf{6\,530.3}   \\ \cline{1-4}
		\multicolumn{1}{|c|}{\textbf{Online Solution Inference}}    & \multicolumn{1}{c|}{0.005}  & \multicolumn{1}{c|}{$\times 4$}   & \textbf{0.02}      \\ \cline{1-4}
	\end{tabular}
	\caption{Computational time (in seconds) for the offline stages of SSM and SGM, and for the online inference stage.}
	\label{tab:CP1}
\end{table}

\begin{table}[H]
	\centering
	\begin{tabular}{|c|c|}
		\cline{1-2}
		Geometry meshing           & 9.0   \\ \cline{1-2}
		Boundary condition setting                 & 0.1   \\ \cline{1-2}
		Semi-discretization       & 0.02   \\ \cline{1-2}
		Time discretization  & 0.1   \\ \cline{1-2}
		\textbf{Total} & \textbf{9.3}   \\ \cline{1-2}
	\end{tabular}
	\caption{Computational time (in seconds) for an implicit FEA solution of the heat transfer problem.}
	\label{tab:CP2}
\end{table}

\section{Conclusions}	
\label{sec:CONC}

This work establishes a foundational framework for a novel approach to generative design (GD), elegantly leveraging optimal transport 
(OT) theory to tackle the challenge of obtaining parametric solutions for engineering problems across new geometrical domains of 
arbitrarily different shapes, particularly when existing solution data for diverse geometries is limited. Recognizing the intricate 
nature of industrial design data, our proposed methodology first employs a re-parameterization of the geometry based on signed 
distance functions to achieve a unified representation of geometrical boundaries. Subsequently, to further unify the representation of
corresponding solutions, our methodology decomposes both the geometry and its associated solutions into Gaussian functions -- a 
technique known as Gaussian splatting in computer graphics -- thereby circumventing the complexities arising from disparate 
geometrical parameterizations and solution dimensions.

To effectively manage the constraints of limited data and the underlying physical principles governing the solutions, we harness the 
power of OT as a robust solution inference tool. To mitigate the data requirements for accurate geometry blending, we strategically 
rely on the notion of Wasserstein barycenters for geometry interpolation. This proves particularly advantageous in shape 
optimization scenarios, allowing for the synthesis of plausible intermediate geometries from a sparse set of examples.

Addressing the inherent computational demands of traditional OT, we introduce a streamlined methodology predicated on discretizing 
fields into a collection of weighted particles, each associated with a Gaussian kernel. This particle-based representation elegantly 
reformulates the OT problem into a computationally tractable optimal assignment problem, drastically reducing its complexity and 
enhancing its practical applicability in design workflows.

The synergistic combination of our OT-based geometry and solution interpolation techniques with classical regression approaches 
empowers the creation of entirely new geometrical domains from existing ones and the subsequent inference of their corresponding 
parametric solutions. The observed error levels in our initial investigations for a parametric, two-dimensional heat transfer problem 
remain remarkably low in comparison to the transported temperature field, underscoring the robustness and accuracy of our approach. 
This compelling performance, coupled with the near real-time responsiveness of the methodology in exploring novel geometrical domains 
and their associated parametric solutions, positions it as a potentially transformative asset within the engineering design process, 
enabling rapid exploration and evaluation of design alternatives.

In conclusion, this research represents a significant initial stride towards a data-driven GD paradigm grounded in the 
principles of OT. Future endeavors will focus on assessing and extending as needed the applicability and robustness of this 
methodology to encompass more intricate and industrially relevant three-dimensional geometries, pushing the boundaries of its 
practical utility. Furthermore, we envision the integration of this framework into inverse problem formulations, thereby empowering 
users to directly generate geometries and their corresponding optimized parametric solutions that precisely meet predefined 
performance requirements. This future direction holds the promise of a truly interactive and goal-oriented design exploration 
experience.

\section*{Acknowledgments}

Sergio Torregrosa, David Munoz, Hector Navarro, and Francisco Chinesta acknowledge partial support from the DesCartes programme (National Research Foundation, Singapore, CREATE) and 
the ESI-KEYSIGHT research chair at ENSAM. Charbel Farhat acknowledges support from the Office of Naval Research under Grant N00014-23-1-2877.
	
\bibliographystyle{unsrt}
\bibliography{biblio.bib}
	
\end{document}